\documentclass{aastex62}

\usepackage{bm}
\usepackage{titlesec}
\usepackage{amsmath}
\usepackage{hyperref}
\usepackage{cleveref}

\newcommand{\s}{\text{s}}

\newcommand{\cm}{\text{cm}}
\newcommand{\km}{\text{km}}

\newcommand{\kpc}{\text{ kpc}}

\newcommand{\sr}{\text{sr}}
\newcommand{\msolar}{\text{$M_{\odot}$}}

\newcommand{\K}{\text{K}}

\newcommand{\erg}{\text{erg}}
\newcommand{\Hz}{\text{Hz}}

\newcommand{\yr}{\text{yr}}

\newcommand{\Mag}{\text{ mag }}
\newcommand{\Mpc}{\text{Mpc}}
\newcommand{\Myr}{\text{ Myr }}

\newcommand{\pd}[2]{\frac{\partial #1}{\partial #2}}

\usepackage{color}
\graphicspath{{./}{figures/}}

\shortauthors{Cain et al.}

\begin{document}

\title{A Model-Insensitive Baryon Acoustic Oscillation Feature in the 21 cm Signal from reionization}

\correspondingauthor{Christopher Cain}
\email{ccain002@ucr.edu}
.
\author[0000-0001-9420-7384]{Christopher Cain}
\affiliation{Department of Physics and Astronomy, University of California, Riverside, CA 92521, USA}

\author{Anson D'Aloisio}
\affiliation{Department of Physics and Astronomy, University of California, Riverside, CA 92521, USA}

\author[0000-0002-5445-461X]{Vid Ir\v{s}i\v{c}}
\affiliation{Kavli Institute for Cosmology, University of Cambridge, Madingley Road, Cambridge CB3 0HA, UK}
\affiliation{Cavendish Laboratory, University of Cambridge, 19 J. J. Thomson Avenue, Cambridge CB3 0HE, UK}
\affiliation{Department of Astronomy, University of Washington, Seattle, WA, 98195}

\author{Matthew McQuinn}
\affiliation{Department of Astronomy, University of Washington, Seattle, WA, 98195}

\author[0000-0001-6778-3861]{Hy Trac}
\affiliation{McWilliams Center for Cosmology, Department of Physics, Carnegie Mellon University, Pittsburgh, PA 15213, USA}


\begin{abstract}

We examine the impact of baryon-dark matter relative velocities on intergalactic small-scale structure and the 21 cm signal during reionization. Streaming velocities reduced clumping in the intergalactic medium (IGM) on mass scales of $\sim 10^4 - 10^8$ M$_{\odot}$. This effect produced a distinct baryon acoustic oscillation (BAO) feature in the 21 cm power spectrum at wave numbers $k\sim 0.1$ h/Mpc, near which forthcoming surveys will be most sensitive. In contrast to the highly uncertain impact of streaming velocities on star formation, the effect on clumping is better constrained because it is set mainly by cosmology and straightforward gas dynamics.  We quantify the latter using coupled radiation-hydrodynamic simulations that capture the Jeans scale of pre-reionization gas.
The clumping factor of ionized gas is reduced by 5-10\% in regions with RMS streaming velocities. 
The suppression peaks $\approx 5$ Myr after a region is reionized, but disappears within 200 Myr due to pressure smoothing. 
We model the corresponding impact on the 21 cm signal and find that the BAO feature is most likely to appear at $\approx$ 10 \% ionization.
During this phase, the feature may appear at the 1 \% (5 \%) level at $k \sim 0.1 (0.06)$ h/Mpc with an amplitude that varies by a factor of $< 10$ across a range of reionization histories. We also provide a model for the signal originating from streaming velocity's impact on ionizing sources, which can vary by 4 orders of magnitude depending on highly uncertain source properties.  We find that the clumping signal probably dominates the source one unless Population III star formation in $10^6 - 10^8$ M$_{\odot}$ halos contributed significantly to the first 10\% of reionization. 

\end{abstract}

\maketitle

\vspace{1cm}

\section{Introduction} 
\label{sec:intro}

The Epoch of reionization (EoR) was the last major phase transition in the Universe, during which the first sources of ionizing photons re-ionized the intergalactic medium (IGM).  In recent years, observational progress has been made towards constraining this epoch. The timing of reionization has been constrained by cosmic microwave background (CMB) optical depth measurements (\citet{Planck2018}). Additional constraints are provided by observations of high-redshift quasars \citep[e.g.][]{Fan2006,Becker2015,McGreer2016,Bosman2018,Davies2018,Eiliers2018,Becker2019} and the population of high-$z$ Lyman-$\alpha$ (Ly$\alpha$) emitters \citep[e.g.][]{Kashikawa2006,Schenker2012,Pentericci2014,Mesigner2015,Inoue2018,Weinberger2019}.    These observations have been effective at constraining timing of reionization, but currently little is known in detail about the reionization process.  Forthcoming observations of the 21 cm spin-flip transition of neutral hydrogen promise a definitive window into the EoR \citep[see e.g.][and references therein]{Furlanetto2006}.  \\

Following a first detection of the EoR 21 cm signal, early efforts will focus on characterizing its brightness temperature fluctuations with the power spectrum.    In this paper, we will investigate whether baryon-dark matter relative velocities (or ``streaming velocities", or just $v_{\rm bc}$; \citet{Tseliakhovich2010}; henceforth T10), which were sourced at recombination, were able to significantly impact the EoR 21 cm power spectrum. Though formally a second-order effect in perturbation theory, $v_{\rm bc}$ was several times the baryon sound speed at decoupling, so its effect on baryonic structure formation is important. 
Previous work has shown that $v_{\rm bc}$ impacts a number of astrophysical processes in the early universe, including star formation in low mass halos \citep{Maio2011,Greif2011,Schauer2019a}, gas content of halos \citep{Naoz2012}, formation of direct-collapse black holes \citep{Tanaka2014}, the BAO feature in the galaxy correlation function~\citep{Blazek2016}, the Ly$\alpha$ forest~\citep{Hirata2018, Givans2020}, and possibly the formation of globular clusters~\citep{Naoz2014, Chiou2019}.   
Of particular relevance for the current paper, $v_{\rm bc}$ has been shown to modify the {\it pre-reionization} 21 cm signal at $z \sim 20$ through its impact on the properties of the first stars and galaxies~\citep[e.g.][]{Dalal2010,Fialkov2012,McQuinn2012,Ali-Haimoud2014,Cohen2016,Munoz2019}.  
These papers have demonstrated that $v_{\rm bc}$ can imprint distinct baryon acoustic oscillation (BAO) features in the 21 cm power spectrum that could be detectable by future experiments.    \\ 

Streaming velocities impacted the universe at two scales that are particularly important for our investigation: (1) near the baryon Jeans scale $k_J \sim 10^2 - 10^3 \text{ h}/\Mpc$, and (2) at the peak of the power spectrum of fluctuations in $v_{\rm bc}^2$, $P_{v^2}(k)$, which occurs at $k \sim 10^{-1} \text{ h}/\Mpc$ (T10, \citet{OLeary2012}).  
The former is roughly the minimum clumping scale of the pre-EoR gas. Recently, \citet{Daloisio2020} (henceforth D20) showed that gas clumpiness on this scale contributes significantly to the ionizing photon budget required to reionize the IGM.  
The suppression of small-scale clumpiness caused by $v_{\rm bc}$, together with any impact $v_{\rm bc}$ has on ionizing photon sources, will therefore translate into fluctuations in the neutral fraction that trace $P_{v^2}(k)$.  
Moreover, near $10^{-1} h/\Mpc$, $P_{v^2}(k)$ is a factor of $\sim 10^2$ larger than the EoR linear matter power spectrum.    
These facts suggest that $v_{\rm bc}$ may have a pronounced effect on the EoR 21 cm power spectrum $P_{21}(k)$ at large scales if it can be written in the form
\begin{equation}
    \label{eq:p21form}
    P_{21}(k) = b_{21,v^2}^2 P_{v^2}(k) + \text{ matter terms}
\end{equation}
where $b_{21,v^2}^2$ is a linear bias factor coupling fluctuations in $v_{\rm bc}$ to fluctuations in the 21 cm signal.  
This $v_{\rm bc}$-sourced term may be detectable in measurements of $P_{21}$ even if $b_{21,v^2}^2 < 10^{-2}$.  Since $P_{v^2}$ exhibits strong BAO features, its appearance in measurements of $P_{21}$ could serve as a ``smoking gun'' signature of reionization. Such a signature would be particularly helpful given the relatively featureless nature of the expected EoR 21-cm power spectrum, and the extreme difficulty of the measurement. Whereas previous studies have explored the coupling of $P_{v^2}$ with $P_{21}$ through the reionization sources~\citep{Bittner2011,Cohen2016}, the effect from gas clumpiness (or the ``sinks") has not been quantified in detail before.   \\

Employing a modified version of the code used in D20, we investigate with fully coupled radiative transfer and hydrodynamics simulations the impact of $v_{\rm bc}$ on the sinks.  We will use our simulation results to quantify their contribution to $b_{21,v^2}$. We will also assess the potential impact of $v_{\rm bc}$ on source properties during the EoR to see whether this effect may contribute at a similar level to sinks.  As we will show, the contribution from the sinks is relatively insensitive to the details of reionization, being mostly fixed by the spectrum of cosmological density fluctuations and gas dynamics.  This will allow us to more tightly bracket the coupling between $P_{v^2}$ and $P_{21}$ from the sinks, as opposed to the highly uncertain effect from the sources.     \\

This paper is organized as follows.  
In \S~\ref{sec:numerical}, we present the details of our simulation code and how we set up our initial conditions.  
In \S~\ref{sec:results}, we present the results of our simulations and describe the impact of $v_{\rm bc}$ during reionization.  
In \S~\ref{sec:21cm}, we model the contribution of sinks to $P_{21}$ analytically and estimate the magnitude of this term using our simulation results.  
We also discuss potential contributions to $b_{21,v^2}$ coming from $v_{\rm bc}$'s impact on source properties and assess the detectability of the $v_{\rm bc}$-sourced signal.  
We summarize in \S~\ref{sec:conc}.  
Throughout this work, we assume the following cosmological parameters: $\Omega_m = 0.305$, $\Omega_{\Lambda} = 1 - \Omega_m$ (flat universe), $\Omega_b = 0.048$, $\sigma_8 = 0.82$, $n_s = 0.9667$, and $h = 0.68$, consistent with the~\cite{Planck2018} results.

\section{Numerical Simulations}
\label{sec:numerical}

We ran a suite of ray tracing radiative transfer (RT) simulations using the same code employed in D20, modified to include $v_{\rm bc}$.   
These are set up to track the response of a highly resolved patch of the IGM to ionizing radiation produced by external sources of constant intensity.
Hence, we do not explicitly model galaxy formation in our simulations. This approach allows us to assess the evolution of the sinks in a controlled manner, whereas it can be difficult to disentangle the physical effects at play in a full simulation of reionization that also models the sources. Our approach also allows us to achieve the required resolution for robustly modeling the clumpiness of the un-relaxed gas (see discussion in D20).

This section describes the code (\S \ref{subsec:code}), initial conditions (\S \ref{subsec:initial}), and simulations included in this work (\S \ref{subsec:simulations}).  

\subsection{The Code}
\label{subsec:code}

We used a modified version of the Eulerian hydrodynamics code of~\citet{Trac2004} that includes the plane-parallel version of the ray-tracing radiative transfer from~\citet{Trac2007}.  
Our simulations were run on one large node of the Bridges supercomputer~\citep{XSEDE} and ran for $200 - 400$ wall clock hours per simulation.  
We assume a gas of primordial composition  with H mass fraction $X = 0.7547$ and He fraction $Y = 1 - X$. The number of hydro cells, RT cells, and dark matter (DM) mass elements are all equal to $N^3$ (with $N = 1024$ in our fiducial runs), and our fiducial box length is $L = 1.024 \Mpc/h$, for a cell length of $1 \kpc/h$.  
The radiation is handled via plane-parallel ray tracing with an adaptive reduced speed-of-light approximation.  Following D20, the radiation sources lie on the boundaries of cubical ``sub-domains" of side length $L_{\rm dom} = 32 \kpc/h$.  
The sub-domain structure of our boxes allows us to ionize all the gas at approximately the same redshift ($z_{\rm re}$) and to maintain a nearly constant photoionization rate throughout the box, which simplifies interpretation of the gas evolution.  
We use a power-law spectrum with intensity $\propto \nu^{-1.5}$ and five frequency bins between 1 and 4 Ryd, roughly typical of the expected energy spectrum of reionization-era galaxies.  All the important heating/cooling processes relevant to primordial gas are tracked by the code after the radiation turns on (see \citet{DAloisio2018}).  
In addition, we keep track of Compton scattering off the CMB at $z > z_{\rm re}$ (which is important for $z \gtrsim 150$) using an approximate analytical fit to the RECFAST free electron fraction\footnote{For $z > z_{\rm re}$, we set $x_e(z) = 0.5 \tanh((z - 1272)/180.6) + 9.309 \times 10^{-5} \times \theta(600 - z) \times z^{0.25} + 0.5$ (where $\theta$ is the Heaviside function) was a good fit to the RECFAST free electron fraction.  }.  
For a more detailed description of the setup of our simulations, we refer the reader to \S 3.1 of D20.  

\subsection{Initial Conditions}
\label{subsec:initial}

We generated Gaussian random field initial conditions at recombination ($z = 1080$), and the time $v_{\rm bc}$ was sourced, using CAMB\footnote{http://camb.info/} transfer functions (TFs). 
We did this to capture the cumulative effect of $v_{\rm bc}$ self-consistently rather than starting from linear theory solutions at lower redshift (as was done in~\citet{OLeary2012} and~\citet{Ahn2018}).  
Following those works, we used separate TFs for Baryons and DM to compute density and velocity growth factors. 
The initial density and velocity fields were generated using the Zel'dovich approximation~\citep[see][for a description]{Padmanabhan1993}.  
We modeled $v_{\rm bc}$ by adding a constant velocity to the gas along the $x$ direction at the initial redshift.  
This approximation is appropriate on scales $\lesssim$ several $\Mpc/h$ because $v_{\rm bc}$ is coherent on those scales (T10). 
We tested the accuracy of our initial conditions prescription by showing that the matter power spectrum produced by our simulations agrees with the linear theory prediction at redshifts when it should (see Appendix~\ref{sec:appB}).
We also compared our results to simulations initialized at a lower redshift to see whether starting from such a high redshift produced spurious shot noise.  
We found (as did~\citet{Hirata2018}) that this was not a significant effect.  

\subsection{Simulations}
\label{subsec:simulations}

Our simulations are run with only hydrodynamics until $z_{\rm re}$. 
At this time, the box is rapidly filled with radiation and all the gas that cannot self-shield is ionized within a few time steps.
The hydrogen photoionization rate $\Gamma_{-12}$ (in units of $10^{-12} \s^{-1}$) at the boundaries of the sub-domains is a free parameter.  We note that $\Gamma_{-12}$ is nearly constant throughout the box due to our sub-domain method.  We ran simulations with $\Gamma_{-12} = 0.3$ and $3.0$, $z_{\rm re} = 6$, $8$, and $12$, $v_{\rm bc} = 20$, $41$, and $65 \km/\s$.  
We used the simulations from D20 with these values of $\Gamma_{-12}$ and $z_{\rm re}$ to allow for comparison to the no-$v_{\rm bc}$ case.  
Note that D20 also considered box-scale density fluctuations by adding a constant background overdensity to some of their simulations.  
We do not do this here because it would make the parameter space unmanageable given the computational cost of our simulations.   
In addition, simulating over-dense regions with nonzero $v_{bc}$ requires a more complex treatment of the initial conditions (see~\citet{Ahn2016, Ahn2018}).  
Simulations with $v_{\rm bc}$ were run down to $z_{\rm stop} = 5$, $5.5$, and $8$ for $z_{\rm re} = 6$, $8$, and $12$, respectively, while the simulations taken from D20 are all run to z = 5.  
Note that throughout this work, the quoted $v_{\rm bc}$ values are those at $z = 1080$, after which $v_{\rm bc}(z) \propto (1 + z)$.  
The values of $v_{\rm bc}$ used here were chosen to facilitate evaluation of integrals of the form
\begin{equation}
    \label{eq:vbc_integral}
    \langle X \rangle_{v_{\rm bc}} = \int_{0}^{\infty} dv_{\rm bc} X(v_{\rm bc}) \mathcal{P}_{v_{\rm bc}} 
\end{equation}
where $X(v_{\rm bc})$ is any quantity of interest and $\mathcal{P}_{v_{\rm bc}}$ is the probability distribution of $v_{\rm bc}$ in the universe, given by~\citep{Tseliakovich2011,Fialkov2014}. 
\begin{equation}
\label{eq:pvbc}
\mathcal{P}_{v_{\rm bc}} = \left(\frac{3}{2 \pi \sigma_{bc}^2}\right)^\frac{3}{2} \times 4 \pi v_{\rm bc}^2 e^{-\frac{3 v_{\rm bc}^2}{2 \sigma_{bc}^2}}
\end{equation}
where $\sigma_{bc} = 30$ km/s is the RMS value.  
Equation~\ref{eq:vbc_integral} is the average of quantity $X$ over the distribution of $v_{\rm bc}$ in the universe.  
Assuming $X(v_{\rm bc})$ can be well approximated by an order $\leq 5$ polynomial in $v_{\rm bc}$, Equation~\ref{eq:vbc_integral} can be evaluated exactly via Gaussian Quadrature with only the three $v_{\rm bc}$ values used here.  \\

Our goal is to capture the impact of $v_{\rm bc}$ on the formation of gas structures at high redshift and quantify how important this effect is once the gas becomes ionized.  
Our simulation setup is well suited to achieve this goal.  
Our simulations have resolution high enough to capture the impact of $v_{\rm bc}$ on the gas at $k \geq 10^2 \Mpc/h$, while being large enough to include structures on mass scales of $10^7 - 10^8 \msolar$, which should be relatively unaffected~\citep{Dalal2010}.  
Hence, it is unlikely that our simulations significantly under or over-estimate $v_{\rm bc}$'s effect on the gas distribution (see Appendix~\ref{sec:appC} for some convergence tests).  
Second, our numerical setup allows us to isolate the effects of $v_{\rm bc}$ on the sinks independently of its effect on sources, allowing for a straightforward interpretation of our results.  
Finally, our use of fully coupled hydro/RT will provide a realistic picture of how $v_{\rm bc}$ ties into the reionization process.  
By modeling the response of the sinks to reionization as in D20, we can make a physically realistic assessment of how important $v_{\rm bc}$ is to their evolution.  

\section{Results}
\label{sec:results}

\subsection{Visualization of the IGM gas structure}
\label{subsec:low_redshift}

\begin{figure}
    \centering
    \includegraphics[scale=0.9]{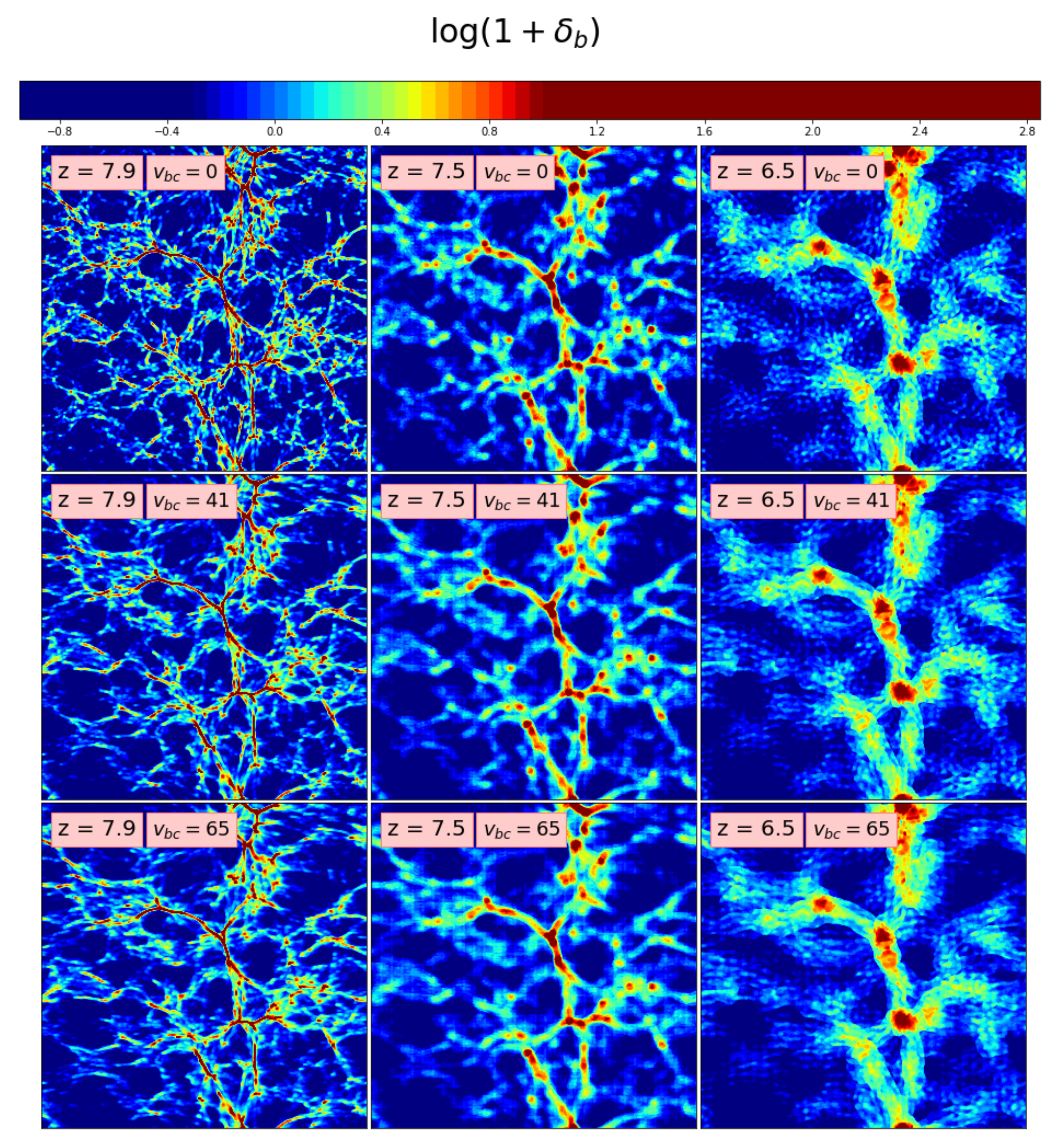}
    \caption{2D slices through the gas density field for redshifts $7.9$, $7.5$, and $6.5$, (left to right) and $v_{\rm bc}$ of $0$, $41$, and $65$ km/s (top to bottom).  The relaxation of the gas is seen going from left to right.  At $z = 7.9$ reduction of small-scale structure by $v_{\rm bc}$ is visible, but after the gas relaxes the differences are too small to easily detect by eye.  We show the results for $z_{\rm re} = 8$ and $\Gamma_{-12} = 0.3$ here; we have checked that they are qualitatively the same for the other combinations of these parameters.  }
    \label{fig:gas_compare_lowz}
\end{figure}

We begin by visualizing the gas structure in runs with different $v_{\rm bc}$. Figure~\ref{fig:gas_compare_lowz} shows 2D slices through the gas density field at redshifts of $7.9$, $7.5$, and $6.5$ (left to right) for $v_{\rm bc} = 0$, 41, and 65 km/s (top to bottom) for $(z_{\rm re}, \Gamma_{-12}) = (8, 0.3)$. After the radiation turns on at $z_{\rm re}$, the gas ionizes quickly, reaching high temperatures. This rapidly increases the pressure in the high density gas filaments, which respond by expanding (``relaxing'') out of their DM potential wells, smoothing the gas density field considerably (see D20 for a detailed discussion).
At redshift $7.9$, most of the gas is still tightly bound in these filaments, but by $z = 6.5$ it has reached the ``relaxed limit'' in which nearly all the small-scale filamentary structure has been erased.  
The relaxation process considerably reduces the clumpiness of the IGM, and with it, the recombination rate.  
This important effect is missed in simulations that do not account for the coupling between hydrodynamics and RT.  \\

At z = 7.9, the impact of $v_{\rm bc}$ is still visible, reflecting the integrated history of the un-relaxed gas.  
However, the differences largely disappear after the gas has relaxed.  
Even by $z = 7.5$, it is difficult to detect by eye any difference between the three runs, and in the relaxed limit at $z = 6.5$ there is no visible difference.  
This result is reasonable, since $v_{\rm bc}$ affects the gas distribution the most on small scales, and it is precisely these scales that are smoothed by the gas relaxation.  
Thus, the majority of the $v_{\rm bc}$ effect does not survive the relaxation process.  
All this suggests that shortly after the gas is ionized, the recombination rate should be appreciably modulated by $v_{\rm bc}$ because of its impact on small-scale structure.  
However, after some time passes the differences should largely disappear owing to the smoothing effect of the relaxation process.   \\

\begin{figure}
    \centering
    \includegraphics[scale=0.9]{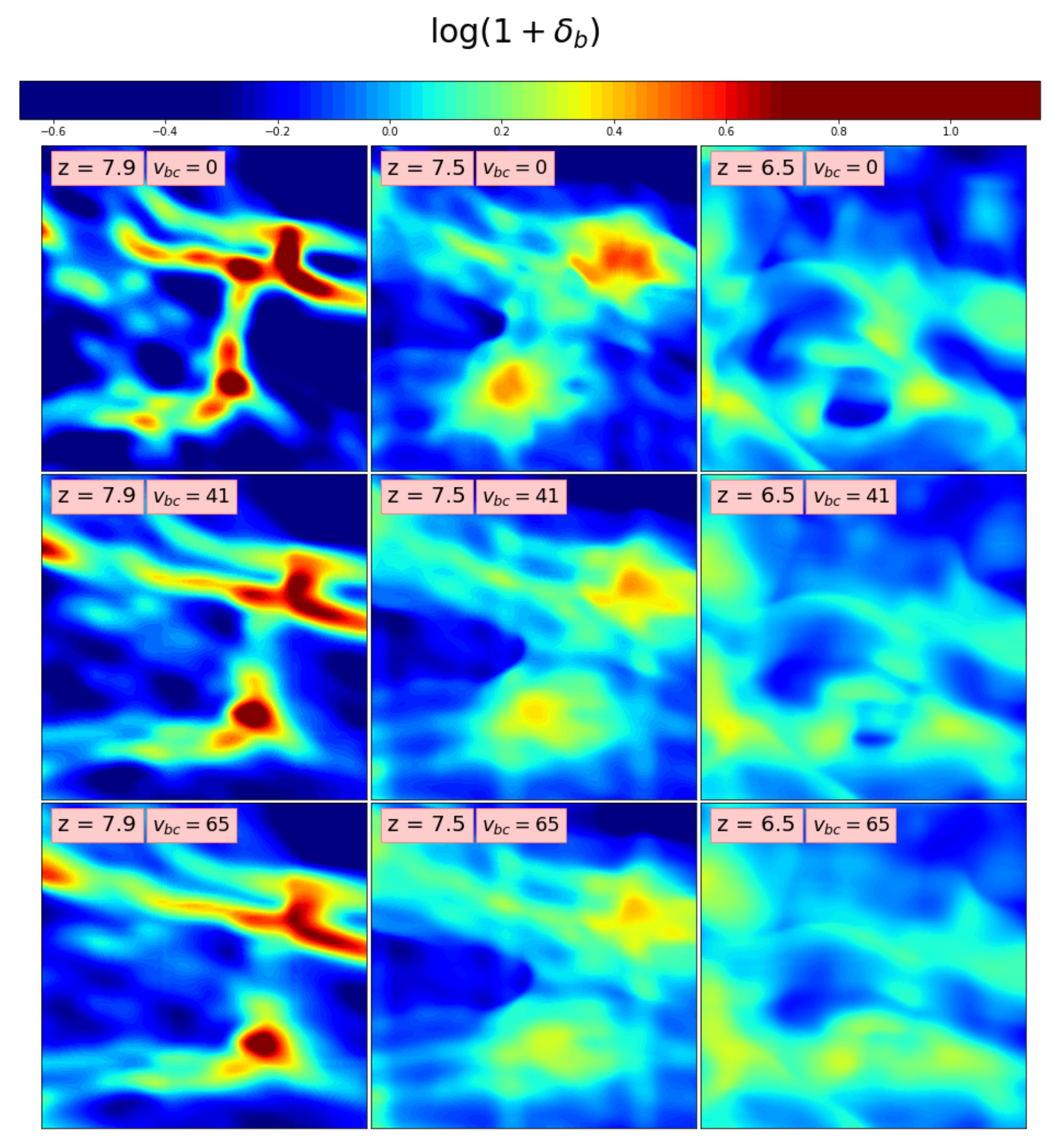}
    \caption{Zoom-in on an $80 \times 80 \kpc^2$ region in Figure~\ref{fig:gas_compare_lowz}.  From left to right, the $v_{\rm bc}$ values are $0$, $41$, and $65$ \km/\s.  The top row shows a marked reduction in structure moving from smaller to higher $v_{\rm bc}$ (left to right).  At the lower redshift, there is much less of a difference }
    \label{fig:gas_zoom_lowz}
\end{figure}

For further clarity, in Figure~\ref{fig:gas_zoom_lowz} we show a $80 \times 80 \kpc^2$ zoom-in of the gas density field at the same redshifts shown in Figure~\ref{fig:gas_compare_lowz}.  
This figure directly compares the initial and relaxed state of the gas.  
At $z = 7.9$, the structures are much more diffuse in the high-$v_{\rm bc}$ runs, and the missing gas fills in some of the voids between structures.  
The effect is less prominent at $z = 7.5$, and almost absent by $z = 6.5$.  
Note also that the structures themselves are different in the relaxed plot.  
This highlights the fact that the relaxation process effectively erases the initial conditions of the un-relaxed gas on small scales, including the $v_{\rm bc}$ effect.  
This relaxation process makes it unlikely that any integrated high-redshift effect that affects only small scales will survive reionization.    

\subsection{Clumping Factor and Mean Free Path}
\label{subsec:clumping}

Based on these results, we expect the recombination rate in a patch of the IGM to be substantially affected by $v_{\rm bc}$ only relatively soon after $z_{\rm re}$.  
Here, we quantify the recombination rate by the clumping factor $C_{\rm R}$, defined to be the ratio of the true recombination rate to that in a uniform-density IGM with constant temperature $T_{\rm ref}$,
\begin{equation}
    \label{eq:clumping_factor}
    C_{\rm R} \equiv \frac{\langle \alpha_B n_e n_{\rm HII} \rangle }{\alpha_B(T_{\rm ref}) \langle n_e \rangle \langle n_{\rm HII} \rangle}
\end{equation}
where $\alpha_B$ is the case B recombination rate for hydrogen, $n_e$ is the free electron density, $n_{\rm HII}$ is the HII number density, and $T_{\rm ref} = 10^4 \K$.
Since all of our simulations have mean densities equal to the global mean and the hydrogen is almost completely ionized after $z_{\rm re}$, we approximate $\langle n_{\rm HII} \rangle \approx n_{\rm H}(z)$ and $\langle n_e \rangle \approx n_{\rm H}(z)(1 + n_{\rm He}(z)/n_{\rm H}(z))$, i.e. assuming singly ionized helium, where $n_{\rm H}(z)$ and $n_{\rm He}$ are the cosmological average number densities of H and He, respectively.  \\

\begin{figure}
    \centering
    \includegraphics[scale=0.85]{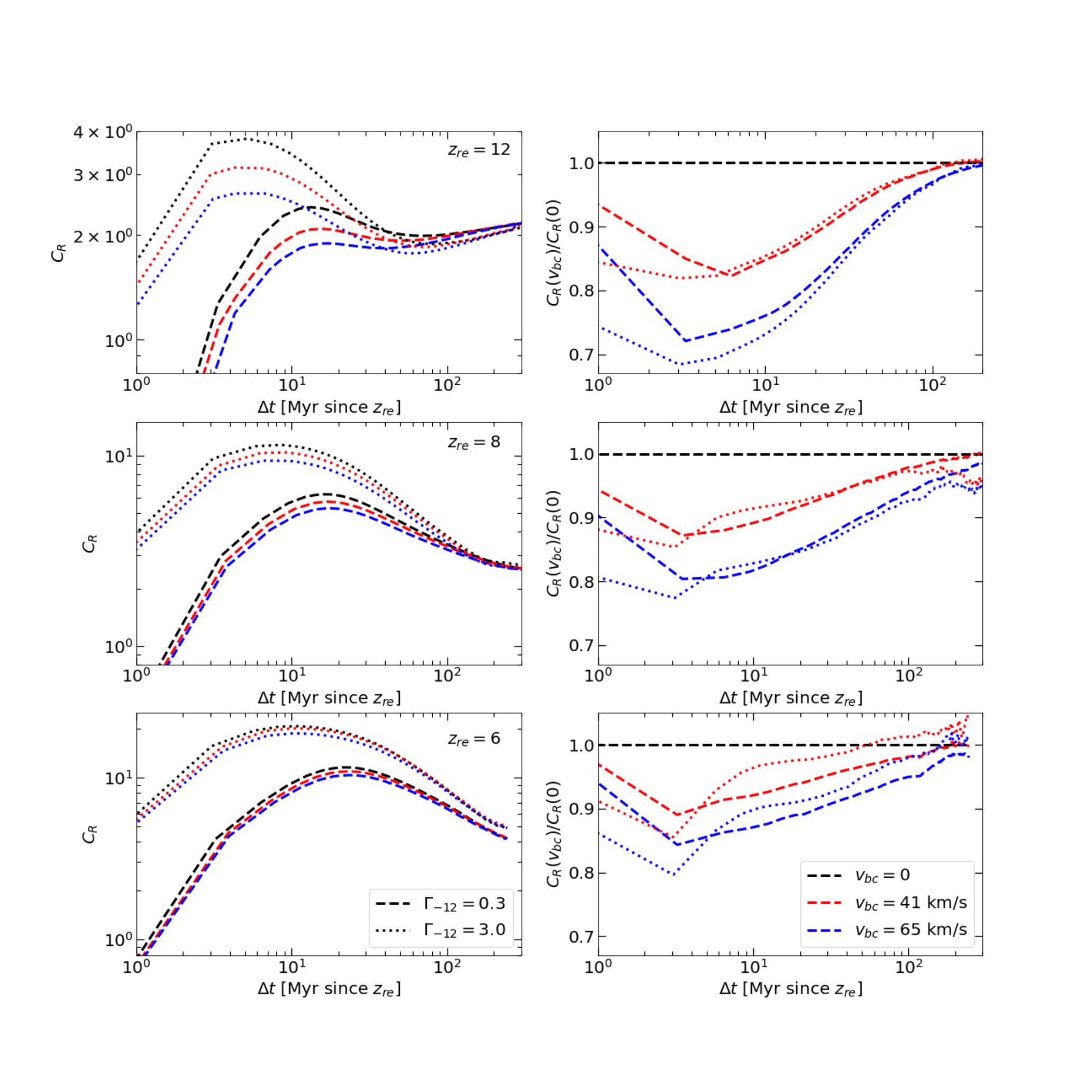}
    \caption{Left: Clumping factor vs. cosmic time since $z_{\rm re}$ = 12 (top), 8 (middle), and 6 (bottom) for $\Gamma_{-12} = 3.0$ (dotted) and $0.3$ (dashed) for all values of $v_{\rm bc}$.
    The difference in $C_{\rm R}$ is largest shortly after the radiation turns on, but attenuates as the gas relaxes.  
    Right: ratio of $C_{\rm R}$ to $C_{\rm R}(v_{\rm bc} = 0)$ for both values of $\Gamma_{-12}$.  
    The difference peaks $5-10$ Myr after $z_{\rm re}$ and steadily declines thereafter, reaching $\sim$ a few percent $200$ $\Myr$ after $z_{\rm re}$. 
    We note that the kinks in the right panels are due to the sparse time stepping at small $\Delta t$.  }
    \label{fig:clump_vs_time}
\end{figure}

We plot $C_{\rm R}$ vs. cosmic time ($\Delta t$) since $z_{\rm re}$ in the left panels of Figure~\ref{fig:clump_vs_time} for $z_{\rm re} = 12$ (top), $8$ (middle) and $6$ (bottom) and $\Gamma_{-12} = 0.3$ (dashed) and $3.0$ (dotted).   
The right panels show the ratio $C_{\rm R}(v_{\rm bc})/C_{\rm R}(v_{\rm bc} = 0)$ i.e. $C_{\rm R}$ as a fraction of the no-$v_{\rm bc}$ case.  
We see that the percentage difference between the different $v_{\rm bc}$ values is largest $\approx 5 - 10 \Myr$ after $z_{\rm re}$, the time at which $C_{\rm R}$ is also at a maximum.  
At this time, the ionizing radiation has penetrated deep into the most overdense regions, but the gas has not yet had time to dynamically relax. 
So, the recombination rate is set by the clumpiness of the initial density field, which is significantly modulated between patches with different $v_{\rm bc}$.  
After $\sim 200 \Myr$, the gas has had time to relax and the fluctuations in $C_{\rm R}$ sourced by $v_{\rm bc}$ have largely disappeared
\footnote{Note that for $\Delta t > 100$ $\Myr$, the clumping factor is actually larger for the $v_{\rm bc} = 41$ case than for $v_{\rm bc} = 0$, particularly in the $z_{re} = 6$ case.  We believe this offset is due to the difference in starting redshift between the $v_{\rm bc} = 0$ simulations and the others.  We tested this by running a set of small box simulations starting at different redshifts, and found that starting at $z = 300$ produces a $\sim 2\%$ suppression in $C_R$ relative to starting at $z = 1080$.  This difference is not large enough to impact our results.  }.  
We emphasize that if our code did not capture the relaxation process, we would significantly over-estimate how much $v_{\rm bc}$ reduces the recombination rate.  
Still, the effect on recently ionized gas is not insignificant, reaching $\sim 15 - 20\%$ for $z_{\rm re} = 12$ and $\sim 10\%$ for $z_{\rm re} = 6$ for $v_{\rm bc} = 41$ km/s.  
Because of the patchy nature of reionization, at any time there will always be some regions in the IGM that were ionized recently and haven't had time to relax.  
In these regions, the recombination rate will depend non-negligibly on $v_{\rm bc}$, potentially leading to detectable fluctuations in the IGM neutral fraction (see the next section).  \\

It has been shown that X-ray heating prior to reionization also reduces the clumpiness of the gas.  D20 ran a simulation in which they set the pre-reionization temperature to a uniform $1000$ K to gauge the maximum effect of X-ray pre-heating. They found that $C_{\rm R}$ was suppressed in a fashion similar to what we find here due to $v_{\rm bc}$.  
This occurs because X-ray preheating raises the pre-EoR Jeans mass, which eliminates structure on the smallest scales.  
In the event that preheating is significant, we expect the importance of $v_{\rm bc}$ to be reduced somewhat as the two processes affect structure at the same mass scales.  \\

\begin{figure}
    \centering
    \includegraphics[scale=0.85]{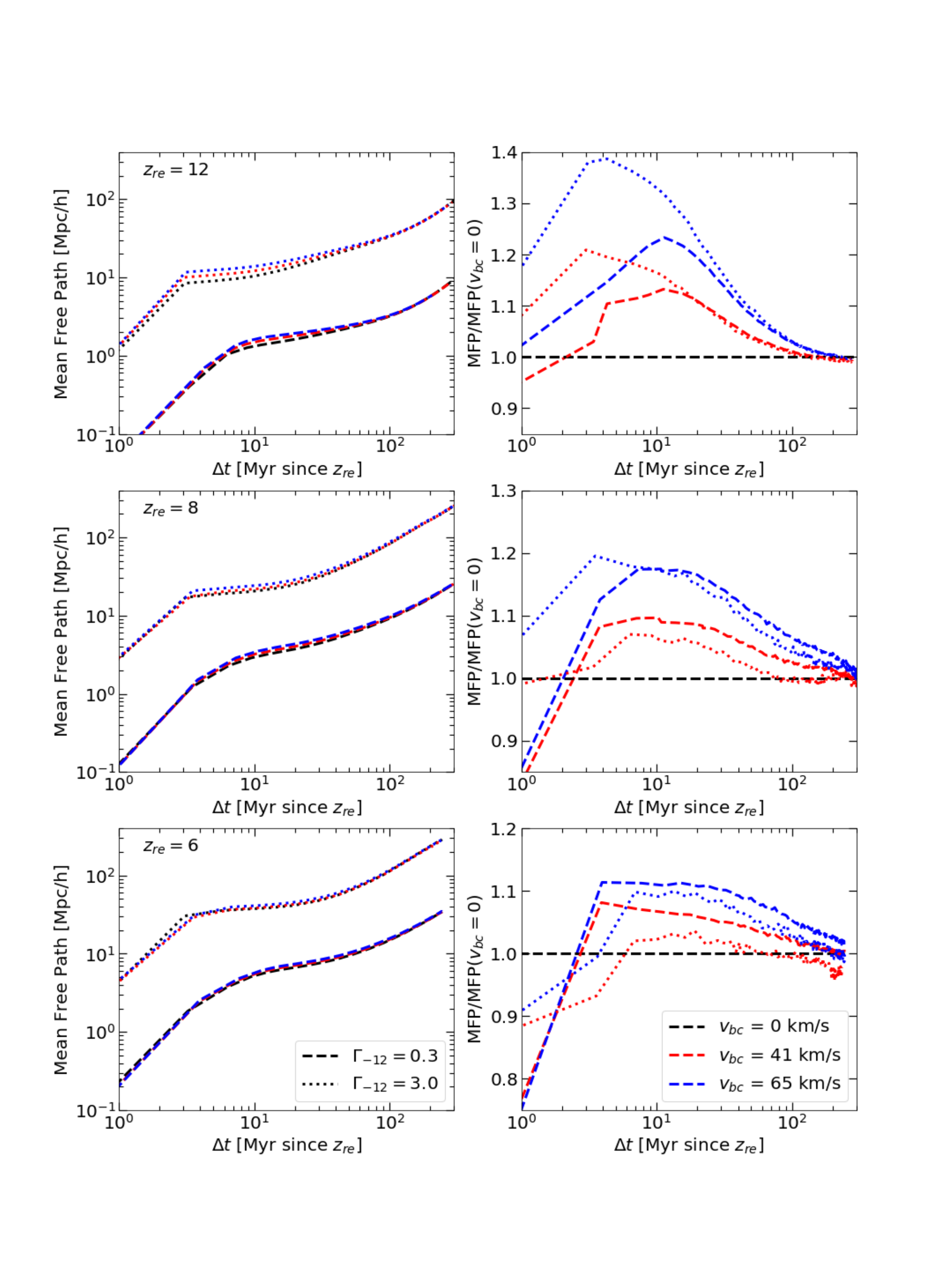}
    \caption{Same as Figure~\ref{fig:clump_vs_time}, but for the MFP of ionizing photons.  The effect of $v_{\rm bc}$ on the MFP is roughly equal and opposite to its effect on the clumping factor.  After $\sim 5-10$ Myr, the difference reaches $25\%$ in the most extreme case, but has largely disappeared after $\sim 200$ Myr. This is much less than the factor of $\sim 10$ difference between the runs with high and low $\Gamma_{-12}$.  }
    \label{fig:mfp}
\end{figure}

Another important quantity during the EoR is the mean free path (MFP) of ionizing photons, which quantifies the typical distance an ionizing photon can travel before being absorbed. 
We calculated the MFP from our simulations using the approach of~\citet{Emberson2013} (see D20 for details).  
Figure~\ref{fig:mfp} shows the MFP from the same simulations as in Figure~\ref{fig:clump_vs_time}.  
We find that $v_{\rm bc}$ modulates the MFP by roughly the same percentage that it does the clumping factor, but in the opposite direction.  
This result is consistent with Figure~\ref{fig:clump_vs_time} because a less clumpy IGM should allow ionizing photons to travel further on average before being absorbed.     
The behavior with time is also qualitatively the same as for $C_{\rm R}$; early on, the MFP is modulated by $10 - 20\%$, but as the gas relaxes the difference disappears.  
Note that the MFP for the $v_{\rm bc}$ runs starts out slightly below the $v_{\rm bc} = 0$ case.  
This is likely because the first regions in the box to ionize are the under-dense ones, which are slightly less dense in the $v_{\rm bc} = 0$ case because more of the gas is locked up in small, dense structures.  
Unlike for $C_{\rm R}$, the percentage difference in the MFP from $v_{\rm bc}$ is small compared to the difference between the runs with high and low $\Gamma_{-12}$.  
Thus, spatial variations in MFP sourced by $v_{\rm bc}$ should be subdominant to those coming from fluctuations in the photoionization rate.  \\

We conclude from these results the IGM recombination rate and MFP are impacted by $v_{\rm bc}$ at the $10-20\%$ level in patches of the universe that have reionized recently. 
After the gas has had time to relax, the effect we observe is largely erased. 
During reionization, there will always be a small percentage of the IGM that was ionized recently, and will therefore display a moderate variation in $C_{\rm R}$ and MFP due to $v_{\rm bc}$.  
These patches will consume fewer ionizing photons due to recombinations than they would in the absence of $v_{\rm bc}$ because of the reduced clumpiness, which will speed up the local reionization process.  
Patches with different values of $v_{\rm bc}$ will be affected differently, leading to fluctuations in the local ionized fraction that trace fluctuations in $v_{\rm bc}$. Granted, these fluctuations should be small, but we also argued in \S 1 that even a tiny coupling between $P_{v^2}$ and $P_{21}$ could result in a detectable BAO feature in the latter. Exploring this possibility is the subject of the next section.   

\section{Impact of $v_{\rm bc}$ on the 21 cm Signal}
\label{sec:21cm}

In this section, we model the impact of $v_{\rm bc}$ on the EoR 21 cm signal and estimate its detectability.  
We begin with some preliminaries in \S \ref{subsec:eor_21cm}.  
In \S \ref{subsec:21cm_fluct}, we adapt the perturbative model for the signal provided by~\citet{McQuinn2018} (henceforth MQ18) to include a $v_{\rm bc}$-dependent term.    
Using this model, we quantify how large the bias factor $b_{21,v^2}$ must be to produce a detectable imprint on $P_{21}$. 
In \S \ref{subsec:bias} and \S \ref{subsec:source_bias}, we assess the contributions to $b_{21,v^2}$ from ionizing photon sinks (\S \ref{subsec:bias}) and sources (\S \ref{subsec:source_bias}).  
Finally, in \S \ref{subsec:detect}, we assess the detectability of the predicted signal using current and future 21 cm experiments.  

\subsection{The EoR 21 cm Signal}
\label{subsec:eor_21cm}

The EoR 21 cm signal is produced by neutral patches of the IGM that have yet to ionize, so its spatial fluctuations set by those of the neutral fraction and the density field~\citep{Furlanetto2006,Pritchard2012}. For simplicity, we neglect redshift-space distortions\footnote{Note that redshift-space distortions have been shown to distort the 21 cm power spectrum at the scales we consider in what follows.  Our conclusions are somewhat dependent on these effects being small.  }~\citep{Jensen2013,Jensen2016} and adopt the typical assumption that Ly$\alpha$ coupling and the first X-ray sources drove the 21 cm spin temperature to be $T_s >> T_{\rm CMB}$ by the time reionization largely begins~\citep{Wouthuysen1952,Madau1997,Pritchard2007}.  Under these assumptions, the 21 cm brightness temperature $T_{21}$ can be written
\begin{equation}
    \label{eq:T21}
    T_{21}(\bm{r},z) = \hat{T}_{21}(z) x_{\rm HI}(\bm{r})(1 + \delta_{\rho}(\bm{r}))
\end{equation}
where $\delta_{\rho}$ is the (nonlinear) matter overdensity, $x_{\rm HI}$ is the neutral hydrogen fraction, and $\hat{T}_{21}$ depends only on cosmological parameters and redshift.  
To first order in over-densities, fluctuations in $T_{21}$ are proportional to $\delta_{\rm HI} + \delta_{\rho}$, where $\delta_{\rm HI}$ is the overdensity in the neutral hydrogen fraction.  
Because the highest-density regions ionized first, $\delta_{\rm HI}$ and $\delta_{\rho}$ will generally have opposite signs early in the EoR \citep[e.g.][]{Giri2019b}.  
For a given wavenumber $k$, the signal will reach a local minimum when $\tilde{\delta}_{\rho}(k) = -\tilde{\delta}_{\rm HI}(k)$ where the tildes denote the Fourier Transform (FT). 
At this time, the dominant density and ionization terms will cancel out and the signal will be sourced entirely by higher-order terms, one of which should be the $v_{\rm bc}$ term in Equation~\ref{eq:p21form}.  
The signal will later reach a local maximum before disappearing entirely when there is no more neutral hydrogen.

\subsection{21 cm Fluctuations}
\label{subsec:21cm_fluct}

Forthcoming surveys will characterize the EoR brightness temperature fluctuations with the power spectrum, defined as $P_{21}(k) \equiv \langle \tilde{\delta}_{21}({k}) \tilde{\delta}_{21}({k}) \rangle$ where $\delta_{21} \equiv x_{\rm HI}(1 + \delta_{\rho})$.  
While modelling the signal is quite complicated and requires numerical simulations~\citep[see e.g.][for a general discussion]{Furlanetto2019a,Koopmans2019,Parsons2019}, 
MQ18 showed that on large scales the power spectrum can be described surprisingly well with perturbation theory.  
They modeled the 21 cm signal using a multi-parameter bias expansion, keeping the minimum number of terms that produced a reasonable fit to the signal in numerical simulations of reionization.  
At large scales and early times, they obtained a good fit using a model with only three parameters; their ``minimal model'' is given by
\begin{equation}
    \label{eq:minimal_model}
    \tilde{\delta}_{21} = b_1 \left(1 - \frac{1}{3}R_{\rm eff}^2 k^2\right)\tilde{\delta_{\rho}} + b_2 \tilde{\delta_{\rho}^2}
\end{equation}
where $\tilde{\delta_{\rho}}$ is the FT of the total matter over-density, and $b_1$, $b_2$ and $R_{\rm eff}$ are time-dependent but scale-independent bias factors.  
In what follows, we will approximate $\delta_{\rho} \approx \delta_1$ in Equation~\ref{eq:minimal_model}, where $\delta_1$ is the linear matter over-density; this approximation is valid at the redshifts and scales considered here.  
$R_{\rm eff}$ roughly characterizes the size of ionized bubbles, which should be small compared to $1/k$ at times and scales considered here, so we will drop it. \\

\begin{figure}
    \centering
    \includegraphics[scale=0.85]{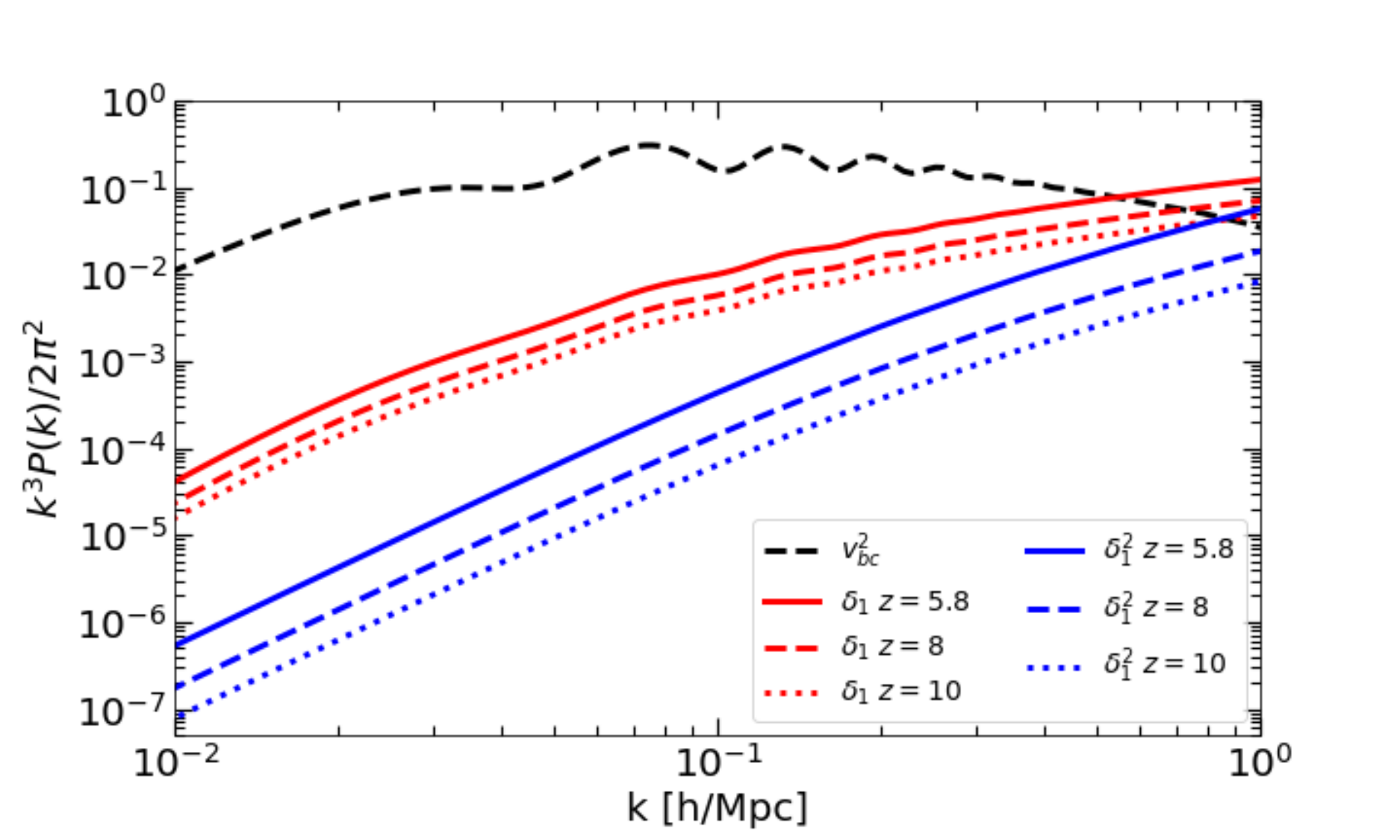}
    \caption{Comparison of the power spectra entering Equation~\ref{eq:p21} at redshifts 5.8 (solid curves), 8.0 (dashed curves), and 10 (dotted curves).  $P_{v^2}$ (black dashed curve) is the same at all three redshifts.  At scales $k \lessapprox 5 \times 10^{-1}$ $h/\Mpc$, $P_{v^2}$ dominates over the matter terms.  }
    \label{fig:pv2_vs_prho}
\end{figure}

We will build upon this model by adding a term proportional to the $v_{\rm bc}^2$ ``overdensity'', $\delta_{v^2} \equiv [v_{\rm bc}^2 - \sigma_{bc}^2]/\sigma_{bc}^2$.   First, we write $x_{\rm HI}$ as
\begin{equation}
    \label{eq:neutral_fraction}
    x_{\rm HI} \equiv \langle x_{\rm HI} \rangle(1 + \delta_{\rm HI})
\end{equation}
where the angle brackets denote an average over the whole IGM.  
Next, we assume that $\delta_{\rm HI}$ is a biased tracer of $\delta_1$, $\delta_1^2$, and $\delta_{v^2}$ and that $\delta_{\rho}$ traces $\delta_{v^2}$.  
Then we have
\begin{equation}
    \label{eq:deltaHI_expansion}
    \delta_{\rm HI} = b_{\rm HI,1} \delta_1 + b_{\rm HI,2} \delta_1^2 + b_{\rm HI,v^2}\delta_{\rm v^2} \hspace{1cm}  \delta_{\rho} = \delta_1 + b_{\rm \rho, v^2} \delta_{\rm v^2}
\end{equation}
where the coefficients are bias parameters.  
Combining Equations~\ref{eq:neutral_fraction} and~\ref{eq:deltaHI_expansion} with the definition of $\delta_{21}$ and dropping all terms 3rd order or higher yields
\begin{equation}
    \label{eq:delta21_expansion}
    \delta_{21} = \langle x_{\rm HI} \rangle (1 + [1 + b_{\rm HI,1}]\delta_1 + b_{\rm HI,2} \delta_1^2 + [b_{\rm HI,v^2} + b_{\rho, v^2}] \delta_{v^2})
\end{equation}
Comparing this to Equation~\ref{eq:minimal_model}, we identify $b_1 = \langle x_{\rm HI} \rangle (1 + b_{\rm HI,1})$ and $b_2 = \langle x_{\rm HI} \rangle b_{\rm HI,2}$ in the case with no $\delta_{v^2}$ term, so we can substitute accordingly to get
\begin{equation}
    \label{eq:delta21}
    \delta_{21} = \langle x_{HI} \rangle + b_1 \delta_{1} + b_2 \delta_{1}^2 + b_{21,v^2} \delta_{v^2}
\end{equation}
where $b_{21,v^2} \equiv \langle x_{HI} \rangle [b_{\rm HI,v^2} + b_{\rho, v^2}]$.
Taking the Fourier transform of both sides of Equation~\ref{eq:delta21} and squaring gives\footnote{Note that the zeroth-order term becomes a delta function at $k = 0$ in Fourier space and thus does not contribute.  }, assuming cross-terms are negligible, 
\begin{equation}
    \label{eq:p21}
    P_{21}(k) = b_{21,v^2}^2 P_{v^2}(k) + b_1^2 P_{1}(k) + b_2^2 P_{2}(k)
\end{equation}
where $P_{1}$ and $P_{2} \propto P_{1} \star P_{1}$ are the first and second order total matter power spectra, respectively, and $P_{v^2}(k)$ is the Fourier transform of $\langle \delta_{v^2}(\bm{x})\delta_{v^2}(\bm{x} + \bm{r})\rangle$.  
\citet{Ali-Haimoud2014} found that the linear and quadratic density fields are uncorrelated on all scales because $\delta_{1}$ $(\delta_{1})^2$ have odd (even) dependence on $v_{bc}$.  
By the same reasoning, $\delta_{1}$ and $\delta_{v^2}$ should be uncorrelated as well.  
Thus our assumption of negligible cross-terms is exact for the terms involving $\delta_{1}$.  
As long as the cross-term between the quadratic terms is sub-dominant to $b_2^2 P_2(k)$, we may safely ignore it when comparing the $v_{bc}$ term to the contribution from the density terms, as we will do shortly\footnote{The neglected cross term will either be featureless and can therefore be absorbed into the $\delta^2$ term or will have BAO features, in which case it may contribute to the signal we are studying.  In either case, we can ignore it as long as it is small.  }.  
Figure~\ref{fig:pv2_vs_prho} plots the dimensionless power spectra ($\Delta^2 \equiv k^3 P(k)/2 \pi^2$) for $\delta_1$, $\delta_1^2$, and $v_{\rm bc}^2$ at redshifts $5.8$, $8$, and $10$ (note that $P_{v^2}$ is independent of redshift).  
For all times shown here, $P_{v^2} > P_{1} > P_{2}$ for $k \lessapprox 5 \times 10^{-1} h/\Mpc$, with the differences growing larger for with increasing redshift and decreasing $k$.  
Moreover, $P_{v^2}$ shows strong baryon acoustic oscillation (BAO) features, suggesting that it's appearance in $P_{21}$ would be distinct even if it only contributes to the total signal at the $\sim 10\%$ level.  \\

\begin{figure}[h!]
    \centering
    \includegraphics[scale=0.7]{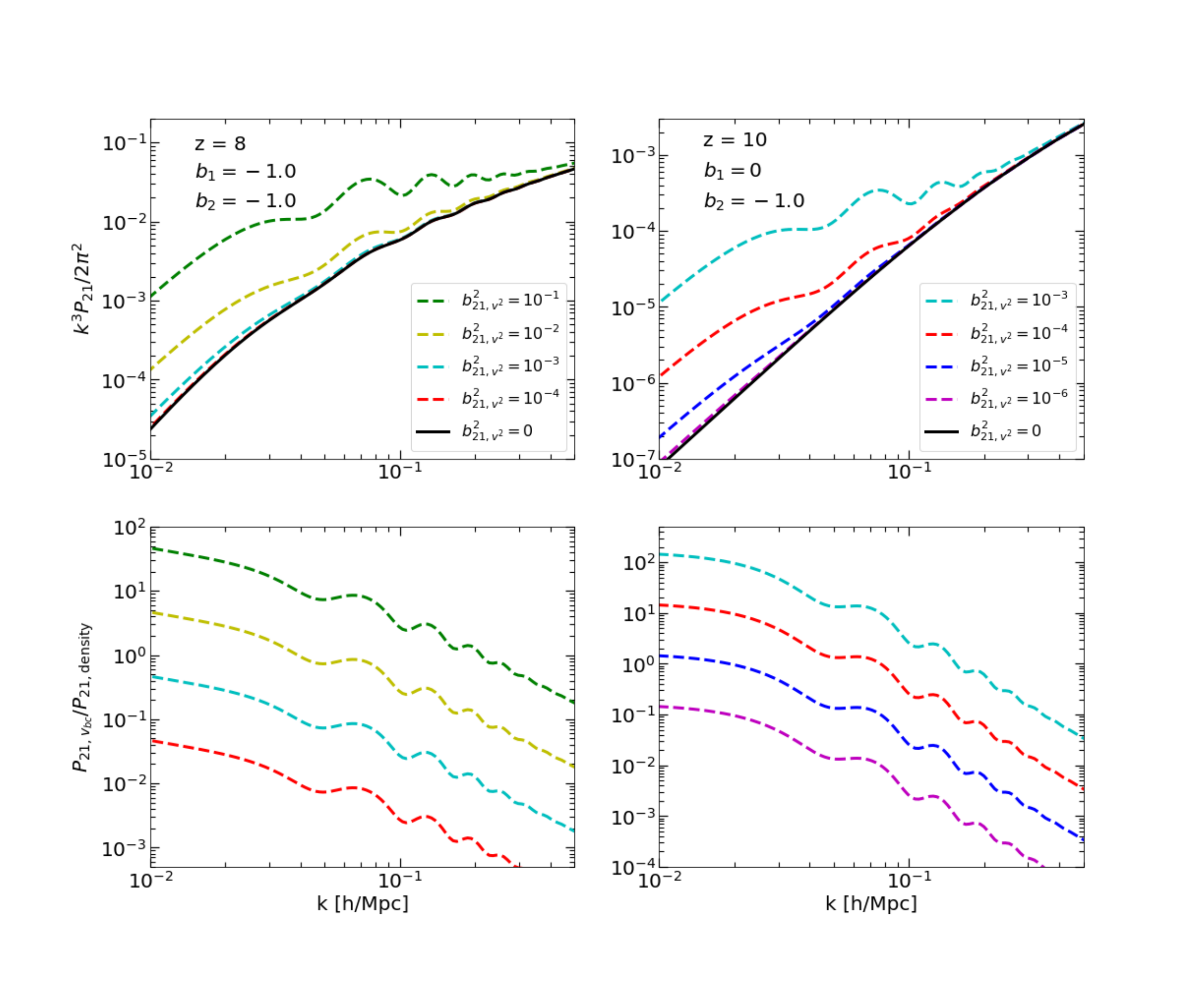}
    \caption{Top Left: Dimensionless 21 cm power spectrum at $z = 8$ for $b_1^2 = b_2^2 = (-1.0)^2$ for $b_{21,v^2}^2 = 0$ (solid black), $10^{-4}$ (red), $10^{-3}$ (cyan), $10^{-2}$ (yellow), and $10^{-1}$ (green).  This plot shows a conservative estimate of $P_{21}$ near its maximum.  Top Right: The same at $z = 10$ for $b_1^2 = 0$, $b_2^2 = 1$, and $b_{21,v^2}^2 = 0$ (solid black), $10^{-6}$ (magenta), $10^{-5}$ (blue), $10^{-4}$ (red), and $10^{-3}$ (cyan).  This plot shows an estimate of $P_{21}$ at its minimum.  The bottom panels show the $v_{\rm bc}$-sourced component as a fraction of the signal without $v_{\rm bc}$ for each of the bias factors.  }
    \label{fig:p21}
\end{figure}

Figure 7 of MQ18 shows how $b_1$ and $b_2$ evolve with time for three idealized models of reionization.  t.
The signal reaches a maximum amplitude when $b_1^2$ is largest (since the $P_1$ term dominates).  
Depending on the model, the maximum value of $b_1^2$ is between $0.5^2$ and $2^2$, at which time $b_2^2 \sim 0 - 3^2$.  
As discussed in~\ref{subsec:eor_21cm}, there is a time early in reionization when $b_1^2 = 0$, and the amplitude of $P_{21}$ is dominated by the second order term, with $b_2^2 \sim 1-2^2$.  
In Figure~\ref{fig:p21}, we show how the $v_{\rm bc}$ term impacts $P_{21}$ in two representative cases for several values of $b_{21,v^2}^2$.   
In the top left panel, we plot the signal at $z = 8$ assuming $b_1^2 = b_2^2 = 1$, representative of the $P_{21}$ maximum, with $b_{21,v^2}^2 \in \{0, 10^{-4}, 10^{-3}, 10^{-2}, 10^{-1}\}$.  
The top right panel shows the same plot at $z = 10$, $b_1^2 = 0$, and $b_2^2 = 1$, representative of the $P_{21}$ minimum, with $b_{21,v^2} \in \{0, 10^{-6}, 10^{-5}, 10^{-4}, 10^{-3}\}$.  
The bottom panels show the range of $b_{21,v^2}^2 P_{v^2}$ as a fraction of the total signal without $v_{\rm bc}$ for the bias parameters considered in the top panel.  
At the $P_{21}$ maximum, $b_{21,v^2}^2 \gtrsim 10^{-3}$ is required to produce a $\sim 10\%$ effect on the signal at $10^{-1}$ Mpc/h, whereas at the $P_{21}$ minimum the same effect is achieved with $b_{21,v^2}^2 \gtrsim 10^{-4}$.  The takeaway here is that $v_{\rm bc}$ has its largest fractional effect at the $P_{21}$ minimum, where the linear contributions from density and ionization cancel, and $P_{21}$ is set by higher order terms in the bias expansion.   \\

We emphasize that the results shown in Figure~\ref{fig:p21} are entirely agnostic about the \emph{cause} of $b_{21,v^2}$. 
In general, we can write
\begin{equation}
    \label{eq:bias_terms}
    b_{21,v^2} = \langle x_{HI} \rangle \left[ b_{\rho,v^2} + b_{HI,v^2}^{\rm sink} + b_{HI,v^2}^{\rm source} \right]
\end{equation}
where the second and third terms come from $v_{\rm bc}$'s impact on the sink and source properties, respectively.  
In what follows, we will make the assumption that $b_{\rho,v^2}$ is essentially $0$, since $v_{\rm bc}$ does not affect the shape of the linear matter power spectrum at scales near $10^{-1} h/\Mpc$ (T10). 
In the next two subsections, we will derive a rigorous model for the contribution to $b_{21,v^2}$ from ionizing photon sinks informed by our simulation results in \S~\ref{sec:results}, and assess analytically the potential contribution from sources.  

\subsection{A Model for $b_{21,v^2}$ from Sinks }
\label{subsec:bias}

In this section, we present our model for the contribution to $b_{21,v^2}$ from the sinks, which we can evaluate using the simulation results presented in \S~\ref{sec:results}. 
Our approach is to relate fluctuations in $v_{\rm bc}^2$ to fluctuations in the ionized fraction $x_i$ via the effect on the clumping factors that we measure from our simulations.  
From this relationship, we will derive an expression for the sinks bias $b_{\rm HI,v^2}^{\rm sink}$ and relate this to the corresponding 21 cm sinks bias $b_{21,v^2}^{\rm sink}$.  \\

We begin with the reionization ``accounting equation'', given by~\citep{Madau1999},
\begin{equation}
    \label{eq:ionizing_accounting_equation}
    \dot{x}_i = \frac{\epsilon}{n_{\rm H}} - \langle C_{\rm R} \rangle \alpha_B n_e x_i
\end{equation}
where $\langle C_{\rm R} \rangle$ is given by
\begin{equation}
    \label{eq:clumping_factor_global}
    \langle C_{\rm R} \rangle(t) = \int_{z_0}^{z(t)} dz_{\rm re}  \mathcal{P}_{z_{\rm re}}(x_i(t)) \int_{0}^{\infty} dv_{\rm bc} \mathcal{P}_{v_{\rm bc}} C_{\rm R}(z_{\rm re},v_{\rm bc},t)
\end{equation}
Here, $\epsilon$ is the emissivity of ionizing photons, $x_i$ is the ionized fraction, and $n_e$, $n_{\rm HII}$, and $n_{\rm H}$ are the mean free electron density, HII number density, and H number density, respectively.  
The integral runs from the start of reionization at $z_0$ to redshift $z(t)$.  
The clumping factor, $C_{\rm R}(z_{\rm re},v_{\rm bc},t)$, which depends on the local values of $v_{\rm bc}$ and $z_{\rm re}$, is extracted directly from our simulations.  
Equation~\ref{eq:pvbc} gives $\mathcal{P}_{v_{\rm bc}}$, and $ \mathcal{P}_{z_{\rm re}}(x_i(t))$ is the probability distribution of $z_{\rm re}$,
\begin{equation}
\label{eq:pzreion}
\mathcal{P}_{z_{\rm re}}(x_i(t)) = \frac{dx_i/dz_{\rm re}}{x_i(t) - x_i(z_0)}
\end{equation}
Note that $\langle C_{\rm R} \rangle$ depends on the ionization history $x_i(t)$ through $ \mathcal{P}_{z_{\rm re}}(x_i(t))$; this is how our model accounts for the ``patchiness'' of reionization.  
However, there are two important dependencies missing from Equation~\ref{eq:clumping_factor_global}: $\Gamma_{-12}$ and the local over-density.  
Because reionization proceeds ``inside-out'' i.e. moves from higher to lower density regions~\citep{Ciardi2003,Furlanetto2004,Mesinger2011} overdense regions are more likely to ionized at higher $z$.  
Because these regions have a higher density of sources, they will also have higher-than average $\Gamma_{-12}$, and the impact of $v_{bc}$ will be different from what we measure here at mean density.  
The top right panel of Figure~\ref{fig:clump_vs_time} shows that the relative $v_{bc}$ effect is slightly larger for higher $\Gamma_{-12}$ at $z_{\rm re} = 12$, with the difference disappearing at lower $z_{\rm re}$.  
Density fluctuation on scales larger than our simulation boxes (so-called ``DC modes'', see~\citet{Gnedin2011}) can be accounted for by re-scaling the local redshift (as was done in D20).  
The effect of increasing the box-scale overdensity should thus be similar to that of lowering $z_{\rm re}$, so we expect a modestly reduced $v_{bc}$ effect in these patches based on Figure~\ref{fig:clump_vs_time}.  
We therefore suggest that the effect early in reionization should be similar to what we calculate here for the mean density case, since higher $\Gamma_{-12}$ and higher density the first patches to ionize drive the effect in opposite directions.  
We also note that X-Ray pre-heating could reduce the effect studied here, since it reduces clumping on the same scales as the stream velocity (see \S~\ref{subsec:clumping}).  
If X-rays and $v_{bc}$ impact the same physical scales (i.e. the Jeans scale), the $v_{bc}$ effect might be substantially reduced by X-rays.  
However, D20 found that pre-heating by X-rays impacts clumping by a factor of 2 as an upper limit, suggesting that the effect on the sinks bias is likely not much more than this.  
A factor of 2 reduction in the sinks bias would result in a factor of 4 reduction in the stream velocity contribution to the 21 cm power spectrum.   \\

We perturb Equation~\ref{eq:ionizing_accounting_equation} by assuming it holds for a spherical patch of the IGM of radius $r$ with mean ionized fraction $x_i^r \equiv \langle x_i \rangle(1 + \delta_{x_i}^r)$ and clumping factor $C_{\rm R}^r = \langle C_{\rm R} \rangle (1 + \delta_{C_{\rm R}}^r)$.  
In doing so, we take Equation~\ref{eq:ionizing_accounting_equation} to be locally true within the patch.  
This is only strictly true if the MFP of ionizing photons is $<< r$.  
Since we are primarily interested in the range $10^{-2} h/\Mpc< k < 10^{-1} h/\Mpc$, we will take $60 \Mpc/h \lesssim r \lesssim 600 \Mpc/h$ for the perturbation scale.  
Figure 11 of D20 shows how the MFP in ionized regions evolves with time for our assumed ionization history (their solid red curve).  
For $\Gamma_{-12} = 0.3$, MFP $<< 60 \Mpc/h$ at $z = 10$ and at $z = 8$ it is still an order of magnitude smaller.  
In the $\Gamma_{-12} = 3.0$ case, the MFP is only a factor of $\sim 5$ less at $z = 10$ and is comparable to $60 \Mpc/h$  at $z = 8$.  
However, since this is the MFP for \emph{ionized} regions only, the MFP with neutral regions included will be considerably smaller, especially early in reionization.    
Moreover, typical values for $\Gamma_{-12}$ extracted from the Ly$\alpha$ forest are in the range $0.3 - 0.5$ with spatial variations by a factor of a few around this value~\citep{Mesinger2009,DAloisio2018,Wu2019}.   
We therefore expect that Equation~\ref{eq:ionizing_accounting_equation} holds locally on the perturbation scales we consider during the majority of the EoR.  \\

Since Equation~\ref{eq:ionizing_accounting_equation} is also satisfied by the IGM mean values $\langle x_i \rangle$ and $\langle C_{\rm R} \rangle$, we can solve for the perturbation $\delta_{x_i}^r(t)$ (see Appendix~\ref{sec:appA} for details).  
Assuming $ \mathcal{P}_{z_{\rm re}}(x_i(t))$ is roughly scale-independent, we may write
\begin{equation}
    \label{eq:ionized_bias_expansion}
    \delta_{x_i}^r(t) = b_{x_i,v^2}(t)\delta_{v^2}^r
\end{equation}
where $\delta_{v^2}^r$ is time-independent and $b_{x_i,v^2}(t)$ is scale-independent.  
An expression for $b_{x_i,v^2}(t)$ can be obtained by Taylor-expanding $\delta_{C_{\rm R}}^r$ to first order in $\delta_{v^2}^r$.  
Since $b_{{\rm HI},v^2} = -\langle x_i \rangle/\langle x_{\rm HI} \rangle b_{x_i,v^2}$ and $b_{21,v^2} = \langle x_{\rm HI} \rangle b_{{\rm HI},v^2}$, we have
\begin{equation}
    \label{eq:b21}
    b_{21,v^2}^{\rm sink} = -\langle x_i \rangle b_{x_i,v^2}
\end{equation}
Hence, we obtain a model for the $v_{\rm bc}$ term in Equation~\ref{eq:p21}.  
The assumption of a scale-independent $ \mathcal{P}_{z_{\rm re}}(x_i(t))$ is valid provided that spatial fluctuations in $ \mathcal{P}_{z_{\rm re}}(x_i(t))$ on the perturbation scale are small compared to the global mean (given by plugging $\langle x_i \rangle$ into Equation~\ref{eq:pzreion}) at each redshift.  
This is not immediately obvious because $\delta_{x_i}^r$ implicitly contains not only the $v_{\rm bc}$ perturbation term, but also matter terms analogous to those in Equation~\ref{eq:delta21}.  
So, it is important to check that $ \mathcal{P}_{z_{\rm re}}(x_i(t))$ is roughly homogeneous on the scales considered here.  
Figure 3 of~\citet{Nasir2019} plots the distribution of $z_{\rm re}$ for three different models of reionization.  
In these plots, the distribution of $z_{\rm re}$ appears to be roughly homogeneous at scales $r \gtrsim 60 \Mpc/h$, justifying our approximation of a scale-independent bias factor. \\

To compute $b_{21,v^2}^{\rm sink}$ from our simulations, we must first solve Equation~\ref{eq:ionizing_accounting_equation} for $\langle x_i \rangle$ and $\langle C_{\rm R} \rangle$ by plugging in our simulation results for $C_{\rm R}(z_{\rm re},v_{\rm bc},t)$ in Equation~\ref{eq:clumping_factor_global}.  
The integral over $v_{\rm bc}$ in this equation can be done via Gaussian quadrature as discussed in~\ref{subsec:simulations}. 
Integrating over $z_{\rm re}$ requires interpolating in two dimensions between the $z_{\rm re} = 12$, $8$, and $6$ $C_{\rm R}$ data as was done in D20 (see their Figure 11).  
To solve Equation~\ref{eq:ionizing_accounting_equation}, we assume the uniform emissivity function from~\citet{Robertson2015} for $\epsilon$ and that $n_e$ and $n_{\rm H}$ assume their cosmological mean values at each redshift.  
Once we have the global history, we can compute the bias (see Appendix~\ref{sec:appA}).  
Figure~\ref{fig:b21v2} shows the results of this exercise for several reionization histories. 
The left panel plots $|b_{21,v^2}^{\rm sink}|^2$ vs. $x_i$ for each history and the right panel plots $x_i$ vs. redshift.
Our fiducial history (red solid curve) starts reionization at redshift $z_0 = 12$ and uses the $C_{\rm R}(z_{\rm re},v_{\rm bc},t)$ from our simulations with $\Gamma_{-12} = 0.3$.  
We also include histories using $C_{\rm R}(z_{\rm re},v_{\rm bc},t)$ from our $\Gamma_{-12} = 3.0$ simulations (``High $\Gamma_{-12}$'', solid cyan), $z_0 = 10$ (Late Start, magenta dashed), an emissivity that is $50\%$ higher (``High Emissivity'', blue dashed).  
Lastly, we include a ``Best Case'' model (green dashed, discussed below) in which the first $\sim 10\%$ of the IGM is ionized almost instantly, and we take $C_{\rm R}(z_{\rm re},v_{\rm bc},t)$ from our $\Gamma_{-12} = 3.0$ simulations.  \\

The bias squared varies in the range $10^{-6} - 10^{-5}$ depending on ionization history near $x_i = 0.1$, but approaches a few times $10^{-5}$ in all histories by $x_i = 0.5$.  
The bias factor is only modestly sensitive to the pace at which reionization proceeds.  
Our fiducial and high $\Gamma_{-12}$ models have the same emissivity function, so they proceed at the same pace early, the latter ending slightly later due to increased recombinations.  
Re-ionization proceeds more quickly in the other three models, ending at the same time as the fiducial model except for the high emissivity case. 
However, the late start and high emissivity models have bias factors that evolve similarly with ionized fraction to the fiducial model.  

The Best-Case model, by construction, provides a rough upper limit on $b_{21,v^2}^{\rm sink}$ at an ionized fraction of $10 - 15\%$, which is around the value of $x_i$ for which $P_{21}$ reaches a minimum, where $v_{\rm bc}$ has its largest fractional effect. 
This is because the ``flash-ionized'' patches reach the time at which the $v_{\rm bc}$ effect is largest coherently, so they contribute maximally to $b_{21,v^2}^{\rm sink}$ all at once.  
However, even in this case the bias squared is only a factor of $\sim 2$ larger than the physically realistic history with the higher $\Gamma_{-12}$ value.  
We therefore do not expect the pace and duration of reionization to significantly impact the sinks bias (although note that we do not consider histories here that begin earlier than z = 12).  
This highlights the relative insensitivity of the sinks bias to details of the reionization history and the properties of the ionizing sources that drove it.  \\

To get a $1\%$ level effect in $P_{21}(z = 8)$ at $k = 10^{-1}$ h/Mpc would require $|b_{21,v^2}^{\rm sink}|^2 \sim 10^{-4}$ (see Figure~\ref{fig:p21}), so for any of these histories the effect would be sub-percent level at the epoch of maximum $P_{21}$.  
Even at $k \sim 10^{-2}$ Mpc/h, where the difference between $P_{v^2}$ and the linear terms is much larger, the effect would still only be a few percent for the bias factors measured here.  
However, at the epoch of minimum $P_{21}$ the results are more promising.  
The Best Case model gives $|b_{21,v^2}^{\rm sink}|^2 \approx 10^{-5}$ at this time, which is enough to change the signal by a few percent at $10^{-1}$ h/Mpc and by $\sim 100\%$ at $10^{-2}$ h/Mpc.  
The other histories (which are physically realistic) give a $\sim 1\%$ effect at $10^{-1}$ h/Mpc and tens of percent at $10^{-2}$ h/Mpc.  
Note that the curves in the left panel of Figure~\ref{fig:b21v2} are very similar (even at low ionized fraction) despite the significant differences in the ionization histories in the right panel.  
This suggests that the sinks bias is constrained to be $|b_{21,v^2}|^2 \sim 10^{-6} - 10^{-5}$ regardless of the details of reionization, e.g. the nature of the source population.   
As we will see in the next section, this is not true of the $v_{\rm bc}$ term coming from the sources themselves.  

\begin{figure}
    \centering
    \includegraphics[scale=0.6]{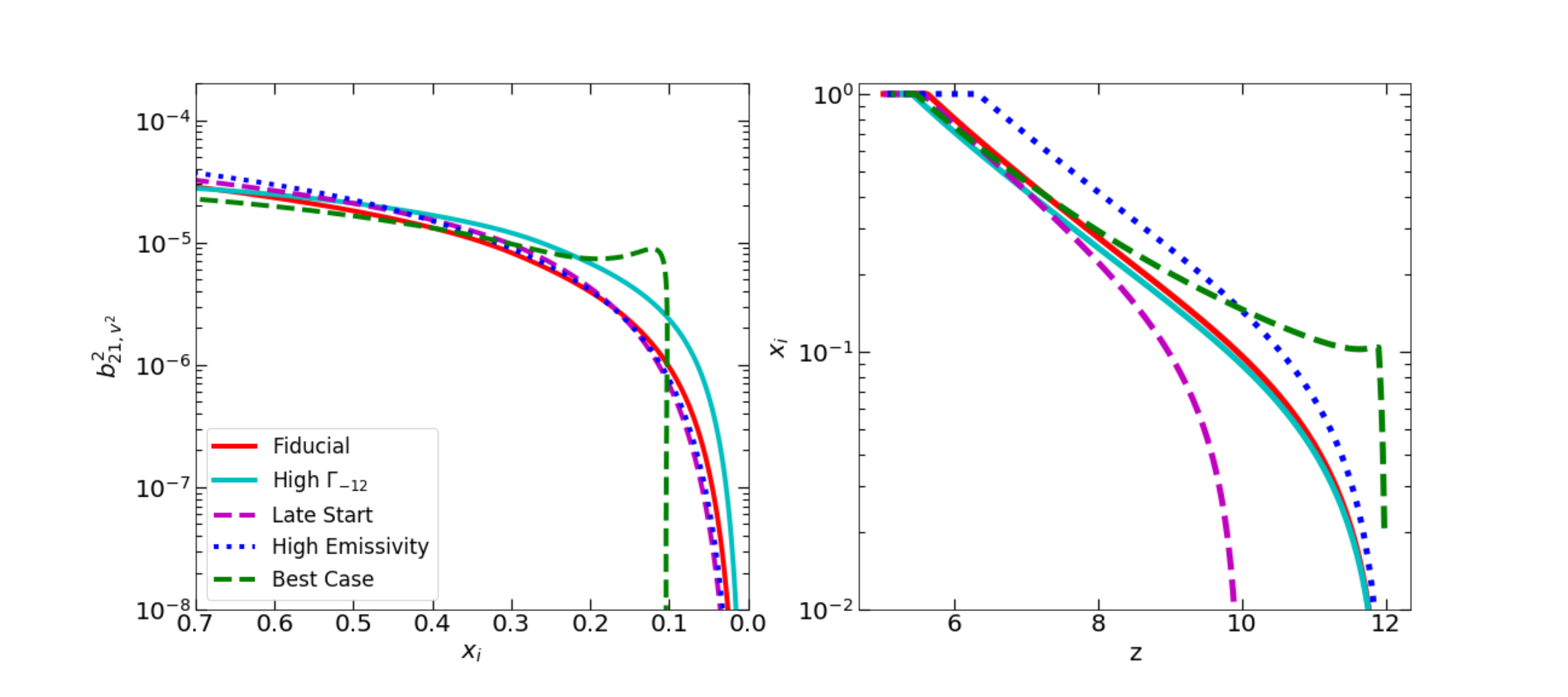}
    \caption{
    Left: $|b_{21,v^2}^{\rm sink}|^2$ vs. ionized fraction for each of the EoR histories discussed in the text. At low ionized fractions, the bias varies in the range $10^{-6} - 10^{-5}$ between the different models, but settles down to a few $\times 10^{-5}$ for all the models late in reionization.  None of these models produce bias factors large enough to produce a detectable signal at the epoch of 21 cm maximum (see Figure~\ref{fig:p21}).  At the 21 cm minimum near $x_i = 0.13$, the Best-Case model gives $|b_{21,v^2}^{\rm sink}|^2 \approx 10^{-5}$, which would alter the minimum signal by a few percent at $k = 10^{-1} h/\Mpc$ and by a factor of $2$ at $k = 10^{-2} h/\Mpc$.  Among the physically realistic histories, the one with high $\Gamma_{-12}$ gives the largest bias.  
    Right: The ionized fraction as a function of redshift for each history.   Note that $\Gamma_{-12}$ changes the ionization history very little early on, but raises $|b_{21,v^2}^{\rm sink}|^2$ by a factor of $\sim 2 - 3$ for low ionized fractions.  
    }
    \label{fig:b21v2}
\end{figure}

\subsection{Contributions to $b_{21,v^2}$ from Sources}
\label{subsec:source_bias}

Here, we discuss possible contributions to the source bias, $b_{21,v^2}^{\rm source}$.  
Since our simulations do not model the sources, we will proceed analytically and use results from the literature where appropriate.  
Previous work has demonstrated that $v_{\rm bc}$ has an important effect on the star formation rate in ``minihalos," with masses $10^6 - 10^8$ M$_{\odot}$, within which it is believed the first stars (Pop III) formed (\citet{Dalal2010},~\citet{McQuinn2012},~\citet{Fialkov2012},~\citet{Munoz2019}, to name a few).  
Primarily, $v_{\rm bc}$ raises the minimum halo mass able to form stars, thereby modulating the radiation backgrounds produced.  
As a result, $v_{\rm bc}$ could modulate the 21 cm spin temperature $T_s$ near the start of reionization. 
 This could change the signal significantly if the assumption $T_s >> T_{\rm CMB}$ is not valid near the start of reionization, which would introduce an extra factor of $1 - T_{\rm CMB}/T_s$ in Equation~\ref{eq:T21}. Additionally, the suppression of ionizing photon production by $v_{\rm bc}$ would work opposite the direction of the sinks by slowing down reionization locally, thereby increasing the 21 cm signal in patches with higher $v_{\rm bc}$.  Although it is widely believed that halos more massive than $10^8$ $M_{\odot}$ (which are less affected by $v_{\rm bc}$) drove reionization, the degree to which Pop III star formation in minihalos contributed to its early phases is highly uncertain. In this section we continue to assume $T_s \gg T_{\rm CMB}$ and we attempt to quantify the coupling of $P_{v^2}$ with $P_{21}$ through the source bias, $b_{21,v^2}^{\rm source}$.      \\

To estimate the impact of $v_{\rm bc}$ on halos, we begin with a general expression for the star formation rate density (SFRD) during reionization (\citet{Sun2016}, \citet{Irsic2019}).  
\begin{equation}
    \label{eq:rho_sfr}
    {\dot \rho}_{\rm SFRD} = \int_{M_{\rm c}(v_{\rm bc})}^\infty dM n(M) f_\star(M) {\dot M}\frac{\Omega_b}{\Omega_m}
\end{equation}
where $n(M)$ is the halo mass function, $\dot{M}$ is the halo accretion rate, and $f_\star(M)$ is the mass-dependent star formation efficiency.   
This integral contains contributions from both Population II and III stars, with the mass cutoff at $M_{\rm atom} \equiv M_{\rm vir}(T_{\rm vir} = 10^4 \K)$, is the atomic cooling threshold, given by inverting Eq. 26 of~\citet{Barkana2001}.
For the lower limit $M_{\rm c}(v_{\rm bc})$, we assume
\begin{equation}
    \label{eq:mc}
    M_{c}(v_{\rm bc},z) =  M_{\rm vir}(500 \K, z) \times \left(\frac{V_{\rm cool}(z,v_{\rm bc},J_{\rm LW}=0)}{V_{\rm cool}(z,v_{\rm bc}=0,J_{\rm LW}=0)}\right)^3 \times [1 + B(4 \pi J_{\rm LW})^\beta]
\end{equation}
where $J_{\rm LW}$ is the specific intensity of the Lyman-Werner (LW) background in units of $10^{-21} \erg/\s/\cm^2/\Hz/\sr$, $(B, \beta) = (7, 0.47)$ for the regular feedback model in~\citet{Munoz2019}, and $V_{\rm cool}(z,v_{\rm bc},J_{\rm LW} = 0)$
is the minimum circular velocity for star formation in the absence of LW feedback, derived from simulations by~\citet{Fialkov2012} (their Eq. 2).  
To obtain $J_{\rm LW}$, we combine Eq. 8 of~\citet{Mebane2018} with Eq. 6 of~\citet{McQuinn2012} and include the LW opacity correction from~\citet{Irsic2019} to obtain
\begin{equation}
    J_{\rm LW} = \frac{7.28}{4 \pi} \times \frac{(1 + z)^3}{H(z)} e^{-\tau_{\rm LW}} (N_{\rm LW}^{\rm II}  \dot{\rho}_{\rm SFR}^{\rm II} + N_{\rm LW}^{\rm III} \dot{\rho}_{\rm SFR}^{\rm III} )
\end{equation}
where we take $N_{\rm LW}^{\rm II} = 9690$, $N_{\rm LW}^{\rm III} = 10^5$, and $e^{-\tau_{\rm LW}} = 0.5$ following~\citet{Irsic2019} and the units of $H(z)$ and SFRD are $\km/\s/\Mpc$ and $\msolar/\yr/\Mpc^3$, respectively.  
For $f_\star(M)$, we used the form in~\citet{Fialkov2014b} for Pop III stars (their Eq. 2) and the form in~\citet{Furlanetto2017b} for Pop II stars (their Eq. 10), where we have tuned the parameters of the latter to give Pop II SFRDs that agree well with the results of~\citet{Visbal2020}.  
The Pop III star formation efficiency at $M_{\rm atom}$, $f_\star^0$, is a free parameter in our model.  
To evaluate Equation~\ref{eq:rho_sfr}, we use the Sheth-Torman mass function and the halo accretion rate given by~\citet{Trac2015} (their Eq. 11) which is calibrated from high-redshift simulations.  
Since $J_{\rm LW}$ and $\dot{\rho}_{\rm SFR}$ are interdependent, we use an iterative scheme to simultaneously solve for them given a value of $f_{\star}^0$.  \\

Following the same formalism as~\citet{Dalal2010} (see their section 2 for details), it can be shown that 
\begin{equation}
    \label{eq:rho_sfr_expansion}
    \dot{\rho}_{\rm SFR} = \langle \dot{\rho}_{\rm SFR} \rangle (1 + b_{\rm SFR, v^2} \delta_{v^2})
\end{equation}
where
\begin{equation}
    \label{eq:rho_sfr_bias}
    b_{\rm SFR, v^2} = -1 + \frac{\langle v_{\rm bc}^2 \dot{\rho}_{\rm SFR} \rangle}{\sigma_{bc}^2 \langle \dot{\rho}_{\rm SFR} \rangle}
\end{equation}
and the averages are over $\mathcal{P}_{v_{\rm bc}}$.  
Depending on the magnitude of $J_{\rm LW}$, $M_{c}$ may be larger or smaller than this cutoff, so we will include contributions from both populations of stars to the bias.  
From here, we can work out an expression for $b_{21,v^2}^{\rm source}$ using the same strategy as in \S \ref{subsec:bias}, but this time by perturbing the emissivity term in Equation~\ref{eq:ionizing_accounting_equation}.  
Using the emissivity model in~\citet{Irsic2019}, we can write these fluctuations as
\begin{equation}
    \label{eq:source_emissivity}
    \frac{\langle \epsilon \rangle}{n_{\rm H}}\delta_{\epsilon} = A_{He} N_{\rm ion}^{\rm III} f_{\rm esc}^{\rm III} \langle \rho_{m}^{-1} {\dot \rho}_{\rm SFR}^{\rm III} \rangle b_{\rm SFR, v^2}^{\rm III} \delta_{v^2}
    \equiv b_{\epsilon,v^2}^{\rm III} \delta_{v^2}
\end{equation}
where $A_{\rm He} = 1.22$, $N_{\rm ion}$ is the number of ionizing photons produced per stellar baryon, $f_{\rm esc}$ is the escape fraction, and $\rho_m = \Omega_m \rho_{\rm crit}(z = 0)$ is the present-day matter density of the universe.  
For Pop III stars we assume $f_{\rm esc} = 0.5$ and $N_{\rm ion} = 40000$.  
We do not include a contribution from Pop II stars to the bias because these stars are expected to form in atomic cooling halos that are unaffected by LW feedback~\citep{Fialkov2012}.  
We therefore do not expect $v_{\rm bc}$ by itself to raise the minimum circular velocity for star formation in these halos above the threshold set by the atomic cooling limit.  
If we ignore the recombination term in Equation~\ref{eq:ionizing_accounting_equation}, the 21 cm bias $b_{21,v^2}^{\rm source}$ is simply the time integral over $b_{\epsilon,v^2}^{\rm III}$.  
We start this integral at $z = 30$, which is early enough to account for the full cumulative impact of Pop III stars on the signal (see~\citet{Munoz2019}).  
Note that this bias factor has the opposite sign of the sinks bias because $v_{\rm bc}$ \emph{reduces} the number of ionizing photons being produced.  \\

\begin{figure}
    \centering
    \includegraphics[scale=0.6]{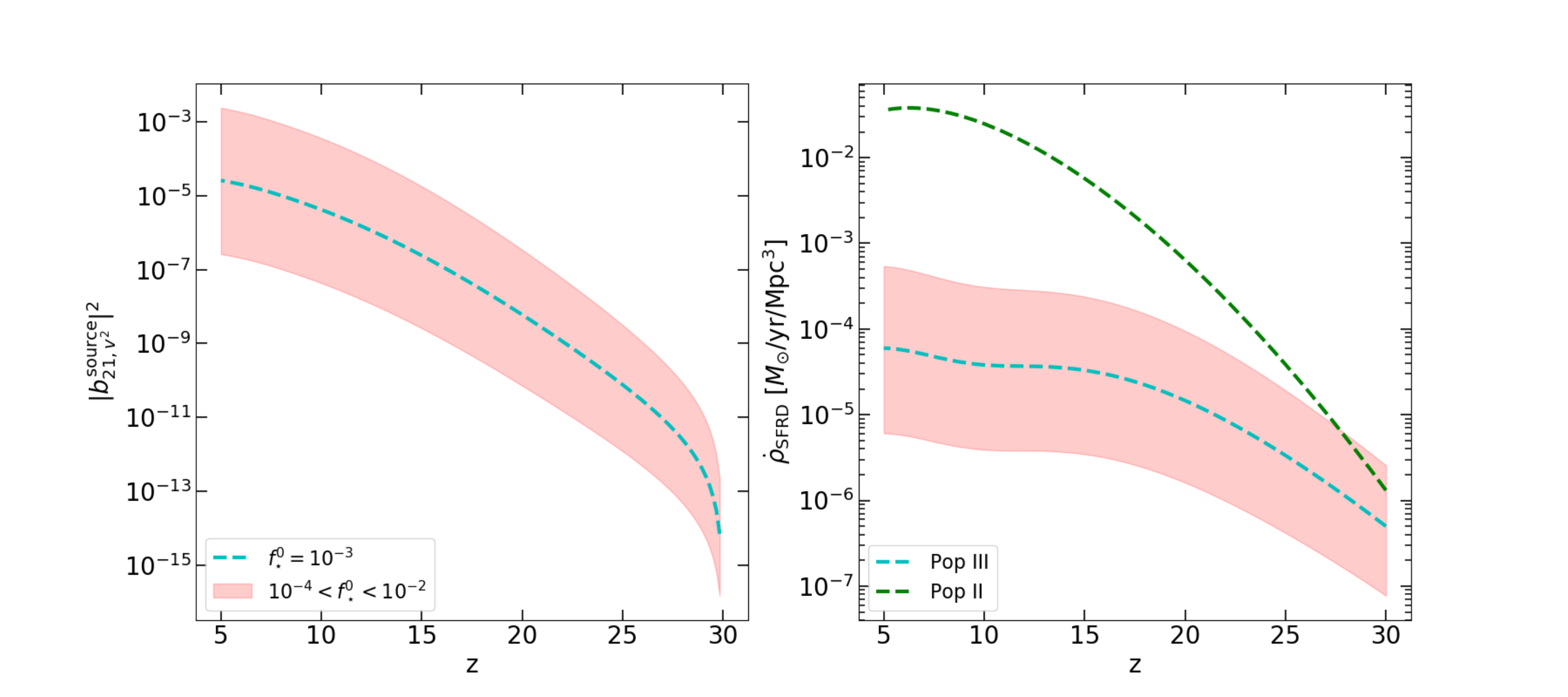}
    \caption{Left: source bias squared for $f_{\star}^{0} =10^{-3}$ (cyan), and the range $10^{-4} < f_\star^0 < 10^{-2}$ (red band). The bias grows with increasing $f_\star^0$, and spans 4 orders of magnitude over the range we consider here for that parameter.  Right: SFRD for pop III and II stars in our model for the same range of $f_\star^0$ shown in the left panel.  The SFRD for both populations agrees well with Fig. 2~\citet{Visbal2020} for $f_\star^0 = 10^{-3}$.  For that choice of $f_\star^0$, our star formation efficiency parameters for both populations are similar to the fiducial values used in that work.  }
    \label{fig:source_bias}
\end{figure}

The left panel of Figure~\ref{fig:source_bias} shows the result of this exercise for $f_\star^0 = 10^{-3}$ (cyan-dashed) and a range of $f_\star^0$ going from $10^{-4}$ to $10^{-2}$ (pink shaded band). 
This range brackets the values generally considered in the literature~\citep[e.g.][]{Trenti2009, Visbal2018, Irsic2019, Visbal2020} as well as other sources of uncertainty (see below).  
At $z = 10$ for $f_\star^0 = 10^{-3}$, the source bias is comparable to the sinks bias ($\sim$ a few times $10^{-6}$) and spans a range of about 2 orders of magnitude above and below this (since $|b_{21,v^2}^{\rm source}|^2 \propto |f_\star^0|^2$).  
The right panel shows the average SFRD for both populations of stars, with the range given for Pop III corresponding to the range of bias factors in the left panel.
The Pop III SFRDs in our model agree reasonably well with those in~\citet{Visbal2020} for the same Pop III star formation efficiencies\footnote{That work assumed a constant Pop III $f_\star$, but because ours depends rather weakly on mass, their $f_{\rm III}$ corresponds closely with our $f_\star^0$}.  
Note that the bias shown in Figure~\ref{fig:source_bias} is formally an upper limit because we neglected the recombination term in Equation~\ref{eq:ionizing_accounting_equation}.  
It is therefore likely that the source bias is less important than the sinks bias early in the EoR\footnote{Note that~\citet{Munoz2019} finds much larger effective bias factors at $z \sim 20$ than we show in Figure~\ref{fig:source_bias}.  However, in that work, the main source of coupling between $v_{bc}$ and $P_{21}$ was the coupling between spin temperature and gas temperature rather than between emissivity and ionized fraction.  We briefly address the possibility of spin temperature fluctuations below.  }.  
However, the result above depends strongly on the assumed values of $f_\star^0$, as well as on how the critical mass $M_c$ is modeled, the assumed accretion rate $\dot{M}$, and the exact relationship between star formation rate and $J_{\rm LW}$.  
For example, weaker LW feedback will result in a larger bias, and a smaller minimum Pop III star formation mass (in the absence of $v_{\rm bc}$ or $J_{\rm LW}$) will increase the bias as Pop III stars will play a larger role in reionization.  
In light of these large uncertainties, we cannot draw definitive conclusions about the magnitude of the EoR source bias.  
Note that these uncertainties highlight the relative precision of our model for the sinks bias.  \\

The case in which $T_s \gg T_{\rm CMB}$ does not hold is even more difficult to assess.  
In this work, we employed the common assumption (e.g. \citet{OLeary2012, Fialkov2014c, Watkinson2015, Giri2019b}) that by the time reionization starts, $T_s$ has been coupled to the gas temperature $T_K$ such that $T_s >> T_{\rm CMB}$, as assumed in Equation~\ref{eq:T21}.  
However, some recent models (e.g.~\citet{Pober2015, Venumadhav2018, Ghara2020}) suggest that this assumption may not hold at the start of reionization.   
This could be either because Ly$\alpha$ photons are inefficient at coupling $T_s$ to the gas temperature, or because heating by X-rays is not efficient enough to raise the gas temperature well above $T_{\rm CMB}$ (see~\citet{Fialkov2014c} for a detailed study).  
In either scenario, to first order Equation~\ref{eq:T21} would be multiplied by a factor $1 - \frac{T_{\rm CMB}}{\langle T_s \rangle} + \frac{T_{\rm CMB}}{\langle T_s \rangle}\delta_{T_s}$.  
The fluctuation $\delta_{T_s}$ would occur because higher $v_{\rm bc}$ would reduce the production of X-Rays and Ly$\alpha$ photons by Pop III stars\footnote{We note that previous work~\citep[e.g.][]{McQuinn2012,Munoz2019} has studied this effect at $15 < z < 30$, but not during reionization.  }.  
This bias would work in the same direction as the sinks bias because it would reduce the amplitude of the 21 cm signal in regions with higher $v_{\rm bc}$.  
For reionization models in which $T_{\rm CMB} = \langle T_s \rangle$ occurs early in reionization, the signal at that time would be dominated to linear order by the $\delta_{T_s}$ term, which may trace $\delta_{v^2}$.  
This may offer another window of time during reionization at which $v_{\rm bc}$ could be detectable.  
Estimating the magnitude of this effect is beyond the scope of this paper, so we leave it to future research. 

\subsection{Detectability of $v_{\rm bc}$}
\label{subsec:detect}

Here we briefly discuss the detectability of the expected signal near the EoR 21 cm minimum (where the fractional effect of $v_{\rm bc}$ is likely to be largest).    
Cosmological 21 cm experiments like the Square Kilometer Array (SKA, \citet{Koopmans2015}) and the Hydrogen Epoch reionization Array (HERA, \citet{DeBoer2016}) are expected to be able to probe the scales discussed here with much higher sensitivity than current experiments, which are struggling to detect the EoR signal at its maximum (such as the Experiment to Detect the Global Epoch of Reionization Signature (EDGES), the Low Frequency Array (LOFAR) and others).  
\citet{Trott2019} show, in their Figure 1, the levels of noise in the dimensionless 21 cm brightness power spectrum $\Delta_{b}^2$ expected for several 21 cm experiments at $z = 8.5$ assuming 1000 hours of integration, including SKA and HERA.  
In the best-case scenario of thermal noise only, they find uncertainties of $\approx 2 \times 10^{-2}$ ($1\times 10^{-2}$) mK$^2$ for SKA (HERA) at the smallest wavenumber that both experiments can detect, $k = 0.06 h/\Mpc$.  
Assuming the thermal noise power spectrum scales as $\nu^{-2 \alpha}$ where $\alpha = 2.55$~\citep{Greig2020}, the thermal noise will be larger by a factor of $\sim 2.1$ at $z = 10$ than at $z = 8.5$.  
The resulting uncertainties are a factor of $\sim 2 (4)$ larger than the density term for SKA (HERA) at this wavenumber; this ratio is roughly the same at $k = 0.1 h/\Mpc$.  
This suggests that the signal near the 21 cm minimum is marginally below the current detection limit of SKA and HERA, assuming these modes are not inaccessible due to foregrounds (see~\citet{Lanman2019} for a discussion of foreground contamination).  
We therefore suggest that future versions of these experiments may be able to detect the EoR 21 cm minimum signal at these wave-numbers if foregrounds can be removed.  \\

\begin{figure}
    \centering
    \includegraphics[scale=0.70]{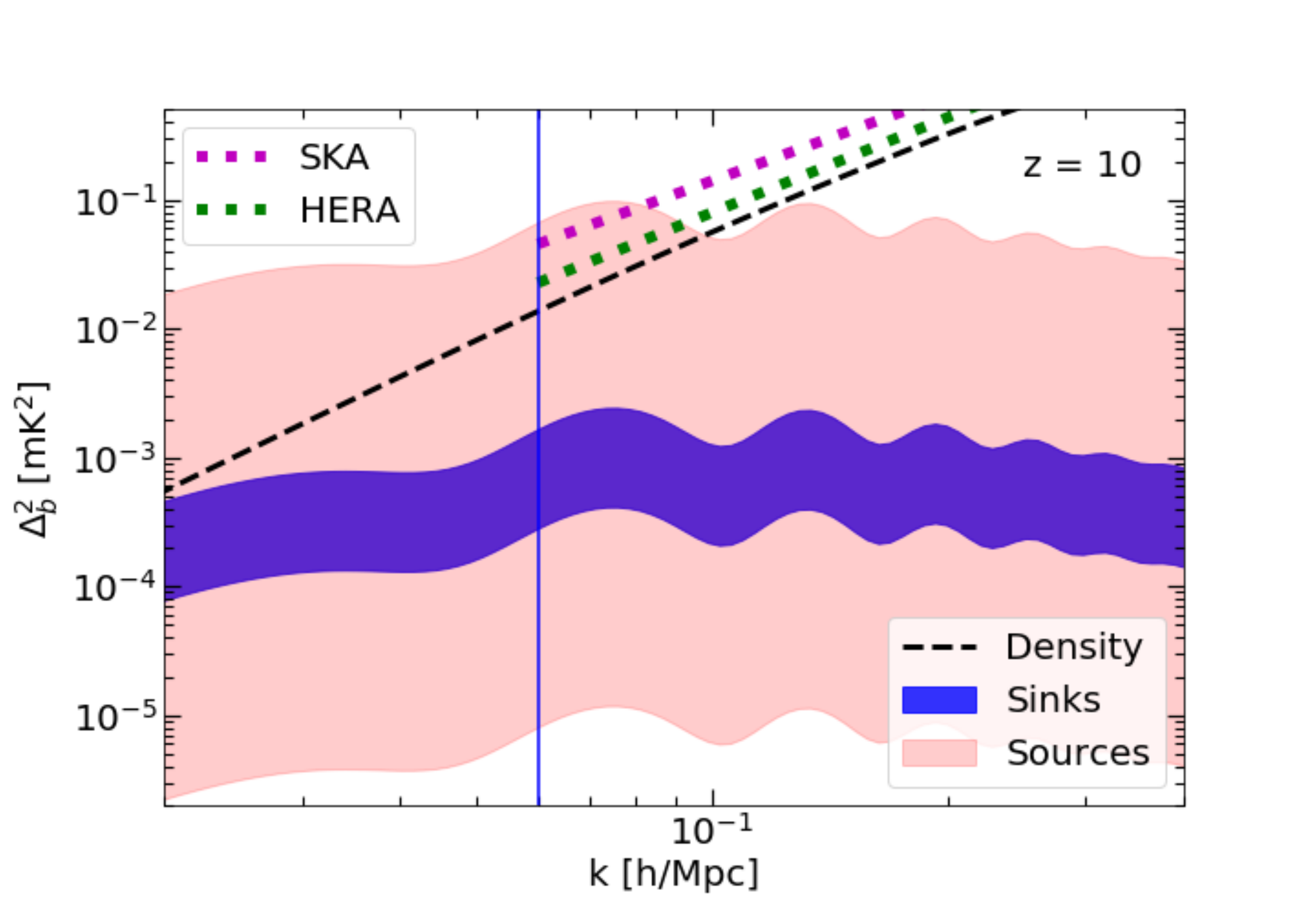}
    \caption{Components of the 21 cm brightness power spectrum at $z = 10$ when the signal is at its minimum during the EoR, using the same density bias parameters as in the upper right panel of Figure~\ref{fig:p21}.  
    We show the contribution from the second-order density term in Equation~\ref{eq:p21} with $b_2^2 = 1$ at $z = 10$ (black dashed).  The blue (red) shaded region denotes the range of possible contributions from the sinks (sources) term discussed above at z = 10.  The blue vertical line denotes $k = 6 \times 10^{-2}$ h/Mpc, and the magenta (green) dashed lines approximately denote the thermal noise limits of SKA (HERA) at and above that wavenumber.  }
    \label{fig:signal}
\end{figure}

Figure~\ref{fig:signal} illustrates the possible contribution of the $v_{\rm bc}$ sourced signal to the total at the EoR 21 cm minimum.  
The black-dashed line denotes the second-order density term in Equation~\ref{eq:p21} with $b_2^2 = 1$ at $z = 10$.  
The blue (red) shaded regions denote the range of signal contributions from the sinks (sources) that we compute in \S~\ref{subsec:bias} (\ref{subsec:source_bias}).  
The blue solid line denotes $k = 0.06$ h/Mpc and the dotted magenta (green) lines roughly denote the thermal noise limits of SKA (HERA) at and above that wavenumber.
At k = $0.06$ h/Mpc, the sinks term contributes $3 - 17\%$ of the signal for an ionized fraction of $13\%$ depending on the reionization history and assumed value of $\Gamma_{-12}$. 
The lower end of this range comes from physically realistic histories with $\Gamma_{-12} = 0.3$, and the high end comes from our Best Case scenario and should be treated as an upper limit.  
A physically realistic history with $\Gamma_{-12} = 3.0$ gives a $7\%$ effect.  
Figure 1 of~\citet{Cohen2016} shows that this epoch of minimum power (which occurs in the range $8 \lesssim z \lesssim 14$ in that paper) should have a duration of at least several tenths of a redshift, which should be a long enough time interval to see the signal if it is detectable.  
As mentioned earlier, a more realistic model would take variations in the photoionization rate and local over-density with $z_{\rm re}$ into account, although it is likely that such an improved model would give the same order of magnitude effect (see discussion in~\S~\ref{subsec:bias}).  \\

Despite these uncertainties, we suggest that a $\sim 5\%$ contribution to the signal from the sinks term is not unrealistic for $k = 0.06$ h/Mpc, provided the linear order term in Equation~\ref{eq:delta21} is close to $0$.
At $k = 0.1$ h/Mpc, the relative contribution of the sinks term is a factor of $\sim 5$ lower than at $0.06$ h/Mpc, so we expect a $\sim 1\%$ contribution at this wavenumber.  
Note that the range of percentages we find for the sinks term varies by only a factor of a few, whereas the source term varies by 4 orders of magnitude for the range of $f_\star^0$ we consider.  
We therefore interpret the sinks term as a lower bound on the $v_{\rm bc}$-sourced signal, except in the very unlikely case that the source and sinks terms happen to exactly cancel each other.  
In addition to the these terms, there will be additional higher-order terms that will achieve their maximum influence at this time as well.  
MQ18 obtained a modestly improved fit to the 21 cm power spectrum using a 7 parameter perturbative model that includes all terms contributing to the power spectrum at 1-loop order (see their Eq. 3.4).  
However, these terms are quite featureless, so although they may contribute to the amplitude of the signal they are unlikely the mask the unique features in the stream velocity term.  
Additional cross terms between the stream velocity term and the higher order matter terms are likely to be much smaller than the stream velocity term.  
Even if these terms are important, they would likely be either featureless (like the matter terms) or contain the same BAO features as the targeted signal (see~\citet{Schmidt2016} for a detailed treatment of similar cross-terms in the context of the low-redshift galaxy power spectrum).  

\section{Summary}
\label{sec:conc}

In this work, we studied the impact of baryon-dark matter relative velocities on the small-scale clumpiness of the IGM during reionization, and how this effect impacts the EoR 21 cm signal. 
Although the streaming velocities were small ($\sim 0.3$ km/s) by the start of reionization, their cumulative effect from earlier times suppressed gas clumpiness, especially in regions where $v_{\rm bc}$ was previously large relative to the sound speed. To quantify these effects, we used high-resolution radiation hydrodynamics simulations that tracked the hydrodynamic response of the IGM to reionization.  
We found that the peak suppression of the clumpiness occurs within the first 5-10 Myr after the gas becomes mostly ionized, before the small-scale structure is erased by Jeans pressure smoothing of the gas.  The clumping factor of ionized gas shows a peak suppression of $5-10\%$ in regions that had streaming velocities of 30 km/s at recombination (approximately the RMS value). 
Differences between regions with and without $v_{\rm bc}$ fall to the percent level by $\Delta t = 300$ Myr, after the gas has had sufficient time to relax in response to the photo-heating from reionization.  \\

To quantify the impact of $v_{\rm bc}$ on the EoR 21 cm power spectrum, we constructed a model for the signal that includes a term coupling $P_{21}$ to fluctuations in $v_{\rm bc}$ through a corresponding bias parameter.  
We modelled contributions to this parameter from ionizing photon sinks and sources.  Using our simulation results for the former, we found that the contribution from sinks is relatively insensitive to the details of reionization, as it is set mainly by the spectrum of primordial density fluctuations and pressure smoothing of the gas.  We found that the characteristic BAO feature imprinted on $P_{21}$ through coupling with the sinks is likely to appear at only the sub-percent level when $P_{21}$ is at its maximum, roughly halfway through reionization.  The feature is most pronounced at $\approx 10 \%$ ionization, when $P_{21}$ is at a minimum.  At this time, the near cancellation of fluctuations in density and ionization allows power from higher-order terms (i.e. from $v_{\rm bc}$) to contribute more significantly.  At the epoch of minimum $P_{21}$, we expect the BAO feature to appear at the $1\%$ ($5\%$) level at $k \sim 0.1$ ($0.06$) h/Mpc due to modulation of the sinks. 
The signal due to sources may be larger than this, but it is subject to a large uncertainty because it depends on poorly-constrained source properties like the star formation efficiency.  
At these wave numbers, the minimum $P_{21}$ that we estimate is close to the thermal noise sensitivity limits of 21 cm experiments like SKA and HERA, so the prospect of detecting the signal in the near future seems low. 
However, it may well be within the capability of the next generation of 21 cm instruments.  

\acknowledgments

We thank the anonymous reviewer for his/her helpful comments on this manuscript.  This research was supported by HST award HST-AR15013.005-A.  This work used the Extreme Science and Engineering Discovery Environment (XSEDE), which is supported by National Science Foundation grant number ACI-1548562. We used the Bridges Large supercomputer at the Pittsburgh Supercomputing Center through allocation TG-AST120066.  We also thank Jonathan Pober and Hyunbae Park for helpful comments on the draft version of this work.  VI acknowledges support by the Kavli Foundation.

\bibliography{references}

\clearpage
\renewcommand{\thefigure}{A\arabic{figure}}
\setcounter{figure}{0}

\appendix

\section{Bias Factor Derivation}
\label{sec:appA}

We provide here the derivation for the sinks' contribution to $b_{x_i,v^2}$ in Equation~\ref{eq:ionized_bias_expansion}.  
Assuming $x_i^r$ and $C_{\rm R}^r$ satisfy Equation~\ref{eq:ionizing_accounting_equation}, we can write a differential equation for $\delta_{x_i}^r$ with the simple form
\begin{equation}
    \label{eq:delta_xi_equation_simplified}
    \dot{\delta}_{x_i}^r + A(t) \delta_{x_i}^r = B(t) \delta_{C_{\rm R}}^r
\end{equation}
where
\begin{eqnarray}
    \centering
    \label{eq:a_of_t_and_b_of_t}
    A(t) \equiv \frac{\dot{\langle x_i \rangle}}{\langle x_i \rangle} + \alpha_B n_e \langle C_{\rm R} \rangle \hspace{2cm}
    B(t) \equiv - \alpha_B n_e \langle C_{\rm R} \rangle
\end{eqnarray}
are functions only of the IGM mean values of $x_i$ and $C_{\rm R}$ for which the solution to Equation~\ref{eq:ionizing_accounting_equation} is known already.  
The fluctuation $\delta_{C_{\rm R}}^r$ is given by the average of local fluctuations over the distribution of $z_{\rm re}$ and $v_{\rm bc}$ within the patch, analogous to Equation~\ref{eq:clumping_factor_global}.  
Assuming the right-hand-side of Equation~\ref{eq:delta_xi_equation_simplified} is a function of time only and not of $\delta_{x_i}^r$ (which will be justified momentarily), the solution is
\begin{equation}
    \label{eq:deltaxi_solution}
    \delta_{x_i}^r(t) = D(t) \int_{t(z_0)}^t dt' F(t') \delta_{C_{\rm R}}^r(t')
\end{equation}
where
\begin{eqnarray}
    \label{eq:d_of_t_and_f_of_t}
    \centering
    D(t) \equiv e^{-\int_{t(z_0)}^{t} dt' A(t')} \hspace{2cm}
    F(t) \equiv B(t') e^{\int_{t(z_0)}^{t} dt' A(t')}
\end{eqnarray}
where $z_0$ is the redshift at which reionization starts.  
Finally, we may write to first order in $\delta_{v^2}^r \equiv (v_{\rm bc}^2 - \sigma_{bc}^2)/\sigma_{bc}^2$,
\begin{equation}
\label{eq:taylor_expansion}
\delta_{C_{\rm R}}^r = \frac{\sigma_{bc}^2}{\langle C_{\rm R} \rangle} \left\langle \pd{C_{\rm R}}{v_{\rm bc}^2}\Big|_{v_{\rm bc} = \sigma_{bc}} \right\rangle \delta_{v^2}^r + \text{ matter terms}
\end{equation}
where the partial derivative is averaged as in Equation~\ref{eq:clumping_factor_global}.  
Equation~\ref{eq:taylor_expansion} is the statement that $v_{\rm bc}$ is a biased tracer of $C_{\rm R}$.  
Combining Equations~\ref{eq:deltaxi_solution} and~\ref{eq:taylor_expansion} yields
\begin{equation}
    \delta_{x_i}^r = b_{x_i,v^2}^r \delta_{v^2}^r
\end{equation}
where $b_{x_i,v^2}^r$ is the scale-dependent ionized fraction bias factor.  
Provided $r$ is large enough that spatial fluctuations in $ \mathcal{P}_{z_{\rm re}}(x_i(t))$ are unimportant, we may write the scale-independent bias factor (Equation~\ref{eq:ionized_bias_expansion}) as
\begin{equation}
    \label{eq:bias_factor}
    b_{x_i,v^2}(t) = \lim_{r \rightarrow \infty} b_{x_i,v^2}^r(t) = \frac{\sigma_{bc}^2 D(t)}{\langle C_{\rm R} \rangle}\int_{t(z_0)}^{t} dt' F(t') S(t')
\end{equation}
where
\begin{equation}
    \label{eq:s_of_t}
    S(t) \equiv \int_{z_0}^{z(t)} dz_{\rm re}  \mathcal{P}_{z_{\rm re}}(x_i(t)) \pd{C_{\rm R}}{v_{\rm bc}^2}\Big|_{v_{\rm bc} = \sigma_{bc}}(z_{\rm re},t)
\end{equation}
Note that $\mathcal{P}_{v_{\rm bc}}$ was absorbed in the definition of $\delta_{v^2}^r$.  

\section{Test of Initial Conditions}
\label{sec:appB}

We tested the initial conditions prescription used in this work by comparing the simulated matter power spectrum at very high redshifts to the expectation from LT.  
We did this primarily to verify that $v_{\rm bc}$ is implemented correctly in our simulations, but also to confirm that starting from $z = 1080$ produces correct results.  
To do this, we ran a set of hydro-only test simulations down to $z = 30$, which are listed in Table~\ref{tab:convergence_sims}.  
These simulations were initialized at $z_{\rm start}$, which is either $1080$ (as in our production runs) or at $z = 300$ (as in D20). \\
	
	\begin{table}
		\begin{center}
			\begin{tabular}{|c|c|c|c|c|c|}
				\hline
				\hline
				Simulation & $z_{\text{\rm init}} $& N & L ($\Mpc/h$) & $v_{\rm bc} (\km/\s)$ \\
				\hline
				Fiducial  & $1080$ & $256^3$ & $0.256$ & $0$\\
				Fiducial + $v_{\rm bc}$ & $1080$ & $256^3$ & $0.256$ & $30$ \\
				High Res & $1080$ & $512^3$ & $0.256$ & $0$ \\
				High Res + $v_{\rm bc}$ & $1080$ & $512^3$ & $0.256$ & $30$ \\
				Low z  & $300$ & $256^3$ & $0.256$ & $0$\\
				\hline
				\hline
			\end{tabular}
		\end{center}
		\caption{List of simulations run to test the accuracy of the initial conditions setup used in this work.  Here, $z_{\text{init}}$ is the initialization redshift of the simulation.  }
		\label{tab:convergence_sims}
	\end{table}

\begin{figure}
		\centering
		\includegraphics[scale=0.85]{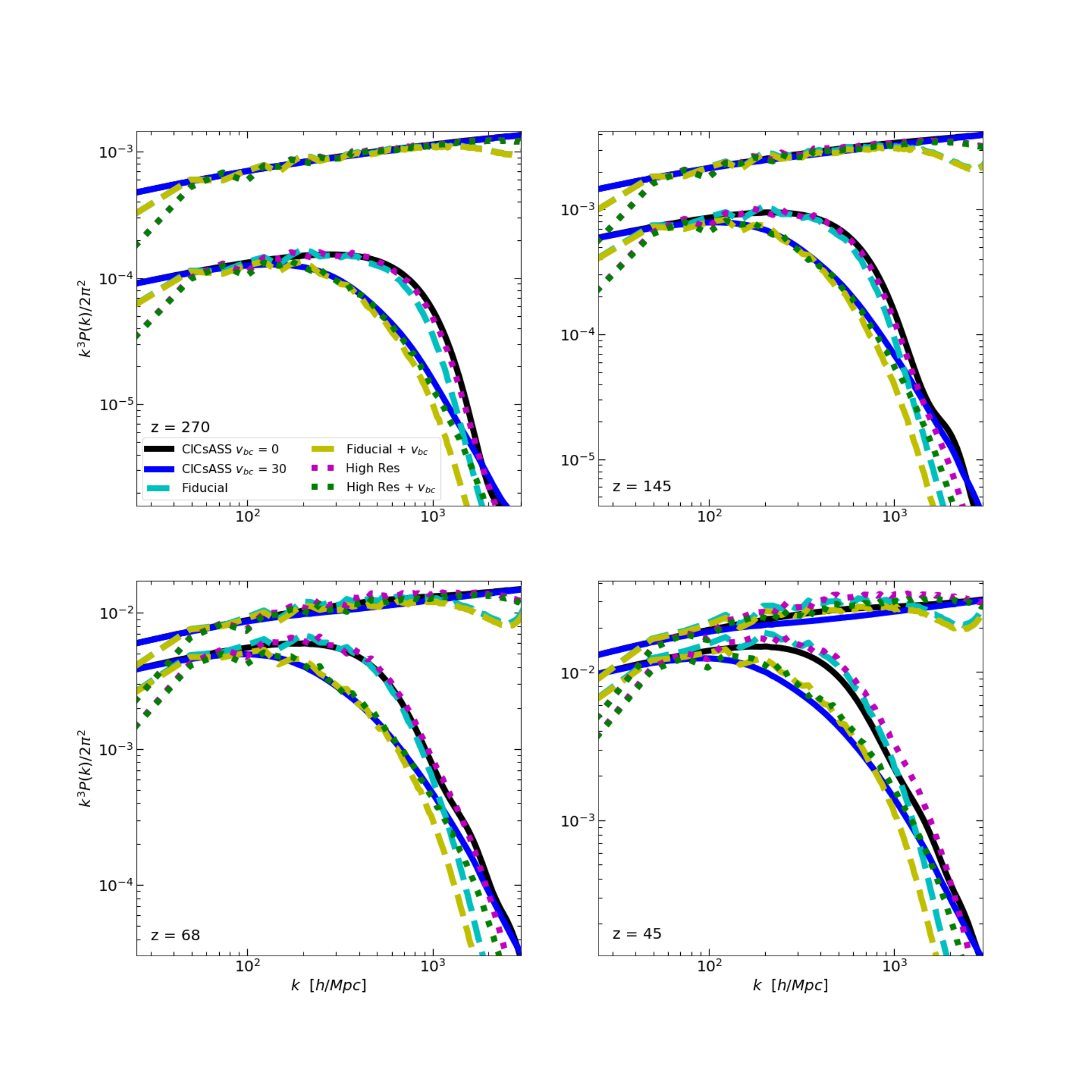}
		\caption{Power spectra of baryons (bottom curves) and DM (top curves) for Fiducial (cyan dashed), Fiducial + $v_{\rm bc}$ (yellow dashed), High Res (magenta dotted), High Res + $v_{\rm bc}$ (greed dotted)  at redshifts $270$, $145$, $68$, and $45$ compared to the LT expectation from CICsASS.  The blue (black) solid curves are the CICsASS LT approximation with(out) $v_{\rm bc}$.  All the DM curves are indistinguishable, and the simulations with and without $v_{\rm bc}$ agree well with their respective LT predictions, especially when the resolution is increased.  }
		\label{fig:res_test}
	\end{figure}

    The matter power spectrum for the tests starting from $z = 1080$ are shown in Figure~\ref{fig:res_test} at redshifts $270$, $145$, $68$, and $45$.  
    The top (bottom) set of curves show the DM (baryon) power spectrum. 
    The solid blue (black) curves are the LT predictions evolved from redshift $1080$ CAMB TFs using the LT approximation from~\citet{OLeary2012} and employed in their initial conditions code CICsASS.  
    Runs with and without $v_{bc}$ have indistinguishable DM power spectra, while the baryon power spectrum is suppressed significantly in the cases with $v_{bc}$.  
    In all cases, the simulations agree well with the LT expectation until $z = 45$ when nonlinear effects begin to become important.  
    The higher resolution runs do a better job at small scales, as expected.  
    Importantly, the simulations with $v_{bc}$ reproduce the CICsASS prediction very well at scales that are captured by the simulations.  
    These results demonstrate that $v_{\rm bc}$ is implemented correctly in our simulations.  \\
	
\begin{figure}
		\centering
		\includegraphics[scale=0.85]{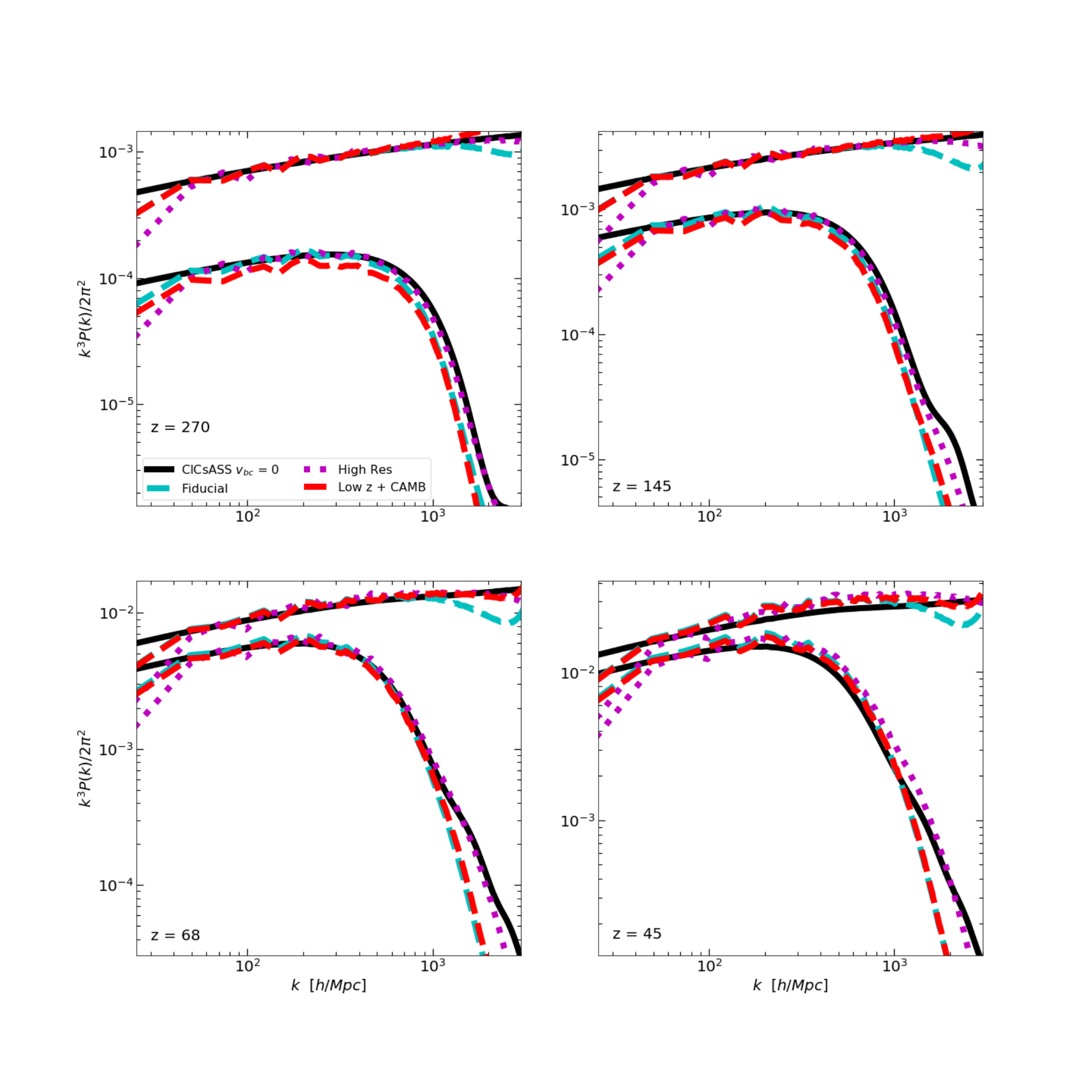}
		\caption{Baryon and CDM power spectra showing how our results vary with different initialization schemes.  The simulations shown are Fiducial (cyan dashed), High Res (magenta dotted), and Low z + CAMB (red dashed).  The Low z run deviates slightly from the others initially, but all three converge at lower redshifts.  This implies that starting from $z = 1080$ and $300$ give very similar results especially at low redshifts.  }
		\label{fig:redshift_test}
	\end{figure}

	We also checked how our results are affected by using different starting redshifts.  
	In Figure~\ref{fig:redshift_test}, we plot the Fiducial (cyan dashed), High res (magenta dotted), Low z(red dashed) power spectra.  
	We compare these to the CICsASS LT expectation without $v_{\rm bc}$ (black solid curve).   
	We see that the simulations initialized at $z = 1080$ (Fiducial and High Res) agree well the LT approximation.  
	The one started from the $z = 300$ CAMB TF deviates slightly from the other two initially.  
	However, after some time has passed, the relative difference decreases, indicating that the slight difference in initial conditions does not affect the results significantly at much later times.  
	This is important for us because it indicates that we can reasonably compare our results to the simulations in D20 (which were initialized in the same was as the Low z run).  
	It also demonstrates that initializing simulations at $z = 1080$ does not introduce significant shot noise, as has been suggested by many previous authors (e.g.~\citet{OLeary2012}).  \\
	
	\section{Effects of Resolution and Box Size}
	\label{sec:appC}
	
	We also assessed the sensitivity of our clumping factor results to numerical resolution and box size.  
	This is important because $v_{\rm bc}$ impacts small-scale gas structures appreciably but leaves the larger structures unaffected.  
	This suggests that too-small boxes would fail to capture the large-scale structures, producing an over-estimate of the $v_{\rm bc}$ effect.  
	Conversely, large boxes with poor resolution would fail to resolve the structures that are most affected, leading to an under-estimate.  
	We quantified these differences by running a set of simulations with $N = 512$ and $L = 256 \kpc$, which gives $8$ times the resolution and $1/64$th the volume of our fiducial runs.  
	We ran simulations with $v_{\rm bc} = 0, 65$ km/s and $\Gamma_{-12} = 0.3, 9.2$ for $z_{\rm re} = 6$.  
	We compared these results to our full box-size run with $v_{\rm bc} = 65$ km/s, $z_{\rm re} = 6$, and $\Gamma_{-12} = 0.3$.  
	We see a maximum suppression in $C_{\rm R}$ relative to the no-$v_{\rm bc}$ case with the same parameters of $\sim 25\% (35\%)$ for $\Gamma_{-12} = 0.3(9.2)$, significantly more than the $15\%$ we got for the fiducial case. 
	We ran a similar set of tests at $z_{\rm re} = 12$, but this time varying the resolution and box size one at a time.  
	We found that increasing box size at fixed resolution reduces the relative $v_{bc}$ effect after about $\Delta t \sim 5 \Myr$, while increased resolution boosts the effect considerably for $\Delta t \lessapprox 10 \Myr$ but not much after this.  
	These results are consistent with the picture that small structures that are affected by $v_{bc}$ dominate the recombination rate early, but after relaxation is complete the recombination rate is set by larger structures that are not appreciably affected by $v_{bc}$.  \\
	
\begin{figure}
	    \centering
	    \includegraphics[scale=1.0]{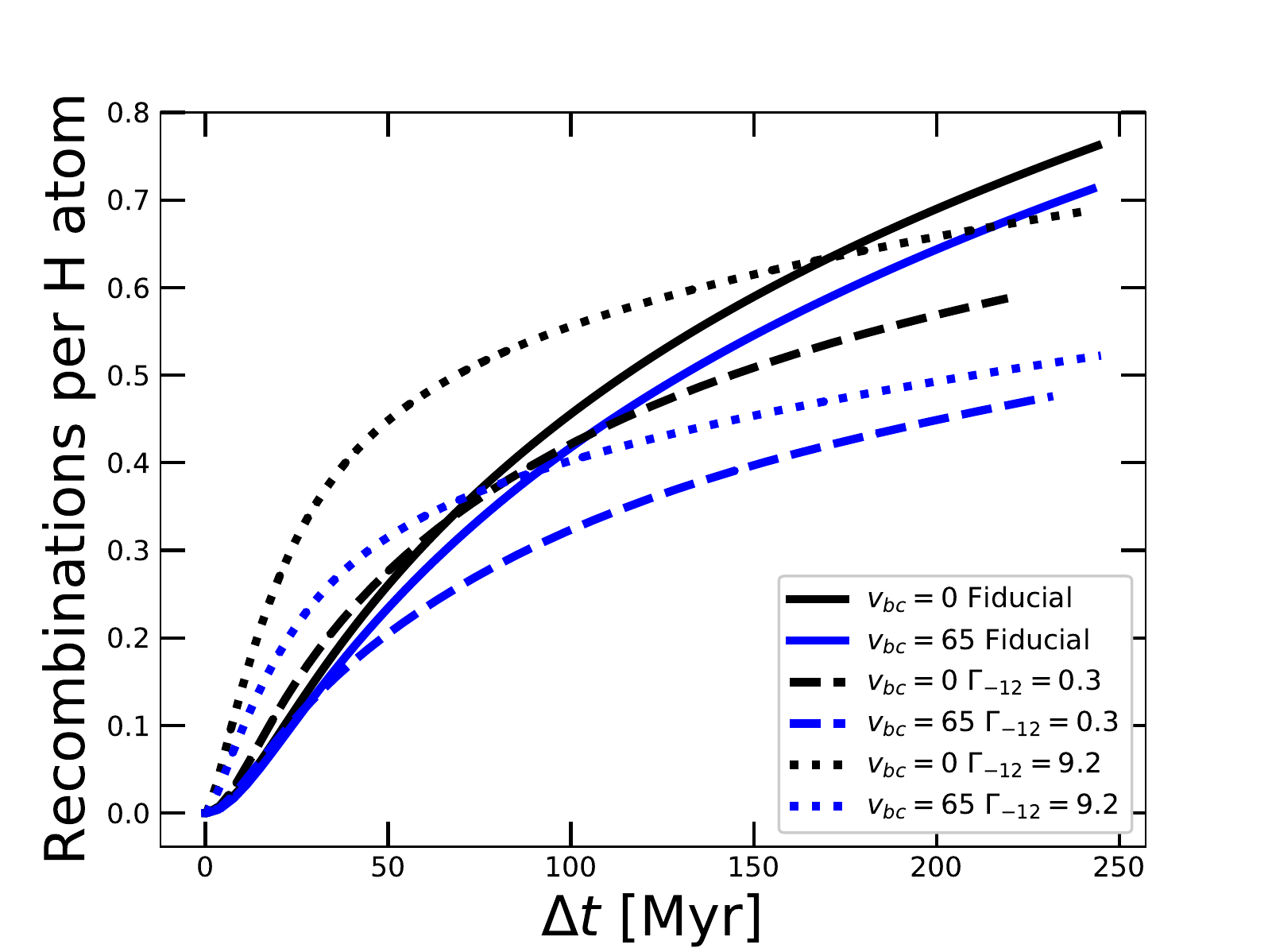}
	    \caption{Number of recombinations per hydrogen atom for our convergence runs (dotted/dashed curves) and our fiducial $z_{\rm re} = 6$, $\Gamma_{-12} = 0.3$ run ((solid curves).  The difference due to $v_{\rm bc}$ is much larger in the smaller boxes, especially the high $\Gamma_{-12}$ case.  This is likely due to a combination of the lack of large systems and better resolution of small ones, both of which enhance the importance of $v_{\rm bc}$.  }
	    \label{fig:recombinations}
	\end{figure}

	In Figure~\ref{fig:recombinations}, we plot the number of hydrogen recombinations per hydrogen atom since $z_{\rm re}$ for the convergence tests at $z_{re} = 6$ alongside our production runs (the fiducial case) with $(z_{re}, \Gamma_{-12}) = (6, 0.3)$, all for $v_{bc} = 0$ (65 km/s).
	The difference between the runs with and without $v_{bc}$ increases for smaller box size/higher resolution and increasing $\Gamma_{-12}$, suggesting that the systems that are resolved in those simulations are more strongly impacted by $v_{\rm bc}$.  
    In addition, the number of recombinations is higher at later times in fiducial case, suggesting that large structures not captured in the smaller simulations contribute a large fraction of the recombinations.  
	This result confirms our suspicion that box sizes that are too small to capture a representative sample of absorbing systems will over-estimate the importance of $v_{\rm bc}$.  
	However, it may be that some of the difference comes from the additional resolution these boxes, in which case our fiducial runs may slightly under-estimate $v_{\rm bc}$'s importance in patches that have been recently ionized.  
	
\end{document}